\documentclass[sn-nature,fleqn]{sn-jnl}

\usepackage{graphicx}%
\usepackage{multirow}%
\usepackage{amsmath,amssymb,amsfonts}%
\usepackage{amsthm}%
\usepackage{mathrsfs}%
\usepackage[title]{appendix}%
\usepackage{xcolor}%
\usepackage{textcomp}%
\usepackage{manyfoot}%
\usepackage{booktabs}%
\usepackage{algorithm}%
\usepackage{algorithmicx}%
\usepackage{algpseudocode}%
\usepackage{listings}%

\usepackage{bm}

\theoremstyle{thmstyleone}%
\theoremstyle{thmstyletwo}%

\theoremstyle{thmstylethree}%

\newcommand{\se}[1]{Section \ref{sec:#1}}

\newcommand{\Fig}[1]{Fig.~\ref{fig:#1}}
\newcommand{\Figs}[1]{Figs.~\ref{fig:#1}}
\newcommand{\Tab}[1]{Table~\ref{tab:#1}}
\newcommand{\be}{\begin{equation}}
\newcommand{\ee}{\end{equation}}
\newcommand{\bad}{\begin{equation} \begin{aligned}}
\newcommand{\ead}{\end{aligned} \end{equation}}
\newcommand{\Msun}{M_\odot}
\newcommand{\Mpc}{\,{\rm Mpc}}
\newcommand{\kpc}{\,{\rm kpc}}

\newcommand{\Gyr}{\,{\rm Gyr}}

\newcommand{\rhobar}{\overline{\rho}}

\newcommand{\tdyn}{t_{\rm dyn}}

\newcommand{\Mv}{M_{\rm vir}}
\newcommand{\Ms}{M_{\star}}

\newcommand{\Mstar}{M_{\star}}

\newcommand{\Rv}{R_{\rm vir}}

\newcommand{\rhalf}{r_{\rm 1/2}}

\newcommand{\Vc}{V_{\rm circ}}

\newcommand{\Vv}{V_{\rm vir}}

\newcommand{\jv}{j_{\rm vir}}

\newcommand{\rmd}{{\rm d}}

\newcommand{\fdisk}{f_{\rm disk}}

\begin{document}

\title[Article Title]{Formation and Environmental Context of Giant Bulgeless Disk Galaxies in the Early Universe: Insights from Cosmological Simulations}


\author*[1]{\fnm{Fangzhou} \sur{Jiang}}\email{fangzhou.jiang@pku.edu.cn}
\equalcont{\footnotesize These authors contributed equally to this work.}
\author[1,2]{\fnm{Jinning} \sur{Liang}}
\equalcont{\footnotesize These authors contributed equally to this work.}
\author[3]{\fnm{Bingcheng} \sur{Jin}}
\author[1]{\fnm{Zeyu} \sur{Gao}}
\author[4]{\fnm{Weichen} \sur{Wang}}
\author[4]{\fnm{Sebastiano} \sur{Cantalupo}}
\author[5]{\fnm{Xuejian} \sur{Shen}}
\author[1,3]{\fnm{Luis C.} \sur{Ho}}
\author[1]{\fnm{Yingjie} \sur{Peng}}
\author[1]{\fnm{Jing} \sur{Wang}}

\affil*[1]{\footnotesize \orgdiv{Kavli Institute for Astronomy and Astrophysics}, \orgname{Peking University}, \orgaddress{\street{5 Yiheyuan Road}, \city{Beijing}, \postcode{100871}, \country{China}}}

\affil[2]{\footnotesize \orgdiv{Institute for Computational Cosmology, Department of Physics}, \orgname{Durham University}, \orgaddress{\street{South Road}, \city{Durham}, \postcode{DH1 3LE}, \country{UK}}}

\affil[3]{\footnotesize \orgdiv{Department of Astronomy}, \orgname{Peking University}, \orgaddress{\street{5 Yiheyuan Road}, \city{Beijing}, \postcode{100871}, \country{China}}}

\affil[4]{\footnotesize \orgdiv{Department of Physics}, \orgname{Universita degli Studi di Milano-Bicocca}, \orgaddress{\street{Piazza della Scienza, 3}, \city{Milano}, \postcode{I-20126}, \country{Italy}}}

\affil[5]{\footnotesize \orgdiv{Department of Physics \& Kavli Institute for Astrophysics and Space Research}, \orgname{Massachusetts Institute of Technology}, \orgaddress{\city{Cambridge}, \postcode{02139}, \state{MA}, \country{USA}}}



\abstract{
Giant bulgeless disk galaxies, theoretically expected to be rare in the early Universe, have been confirmed by the James Webb Space Telescope (JWST) to exist as early as 2 billion years after the Big Bang. 
These morphologically extreme systems offer valuable insights into the physics of disk formation and the interplay between galaxies and their dark-matter halos. 
Using cosmological simulations, we identify analogs of such galaxies with stellar masses around $10^{11} \Msun$ and half-light radii up to 6 kpc at $z \sim 3$ and characterize the factors that contribute to their formation. 
These galaxies form in young cosmic knots, populating host halos of high spin, low concentration, and spherical shapes. 
They feature dynamically coherent circum-galactic medium, as well as gas-rich, coherent mergers, which preserve their disk morphology and drive their large sizes. 
Interestingly, all the simulated giant disks harbor a compact, aligned inner disk, marginally resolvable in JWST images with a S\'ersic index near unity. 
These findings highlight the environmental and structural conditions necessary for forming and sustaining giant bulgeless disks and provide a theoretical framework for interpreting JWST observations of extreme disk morphologies in the early Universe.
}

\maketitle

\section{Introduction}\label{sec: intro}

The study of galactic morphologies provides critical insights into the evolution of the Universe, bridging the connections between luminous matter and the dark-matter halos that host them. 
Among these, disk galaxies stand out as key laboratories for understanding galaxy-halo connections, as they are theoretically expected to contain the structural information of their dark-matter habitat \citep{MMW98}.
However, reproducing disk galaxies in cosmological models has posed significant challenges due to limitations in resolution and the complexities of baryonic feedback processes \citep{Katz92, Navarro97, Weil98}. 
Even today, despite decades of effort, achieving accurate representations of disk formation at the correct epochs continues to be a major hurdle \citep{Vega-Ferrero24}, particularly as the James Webb Space Telescope (JWST) reveals a plethora of disky systems in the early Universe \citep{Huertas-Company24}. 
Modern cosmological models suggest that stable, extended disks were rare in the early Universe due to rapid mass assembly, frequent spin flips on short timescales \citep{Dekel20}, and intense, bursty star formation \citep{Yu21, Hafen22}, all of which hindered the stabilization of disks. 
Numerical studies indicate that high-redshift galaxies typically underwent a phase of concentrated star formation, referred to as {\it gas-rich compaction} \citep{DB14,Zolotov15}.
This process leads to the formation of a stellar bulge that stabilizes subsequently accreted gas, ultimately facilitating disk settlement \citep{Dekel20,Hopkins23}. 

In this prevailing theoretical framework that operates in the early Universe, a stellar bulge is considered a prerequisite for stable disk formation, with disk morphology becoming common after this compact star-forming phase, typically occurring at a characteristic halo mass scale of $\sim10^{11.5}\Msun$ or stellar mass of $\sim 10^{10}\Msun$ \citep{Lapiner23}. 
In contrast to these expectations, JWST recently detected an extraordinary disk galaxy at $z=3.25$, known as the ``Big Wheel" (BW) \citep{Wang24}. 
This system is almost bulgeless, yet it has a half-light radius of $\sim 9\kpc$ and a stellar mass of $\sim 2\times 10^{11}\Msun$.
 The disk size is approximately $3\sigma$ above the size-mass relation at that time \citep{Ward24, Costantin23}.
 
This high-$z$ {\it giant bulgeless disk} (GBD) is fundamentally different from nearby bulgeless giant disks of similar mass and size \citep{Kormendy10}, because they are expected to form in drastically different environments.
The nearby pure disks are hosted by dark-matter halos of masses that are quite common at present day, such that they can reside in low-density cosmic environments and thus have quiescent merging histories \citep{Martig21}. 
High-$z$ giant disks, while having similar masses, inevitably form in high density peaks because of the hierarchical nature of structure formation. 
The high-density environments, however, are exactly where the compaction processes are expected to arise, abundant in mergers and counter-rotating cosmic filaments that can cause angular-momentum loss \citep{Lapiner23}. 
In fact, another giant spiral galaxy at $z\sim3$ of similar size and mass to the BW has a prominent bulge \citep{Umehata24}, in line with the theoretical picture. 
This raises key questions: 
What conditions enable extended disks to form in the early Universe? 
How can the disk be so well-developed in the BW despite the absence of a significant stellar bulge?
These puzzles can be addressed if such GBDs at high $z$ are reproduced in cosmological simulations. 

\begin{figure*}
    \centering
    \includegraphics[width=\linewidth]{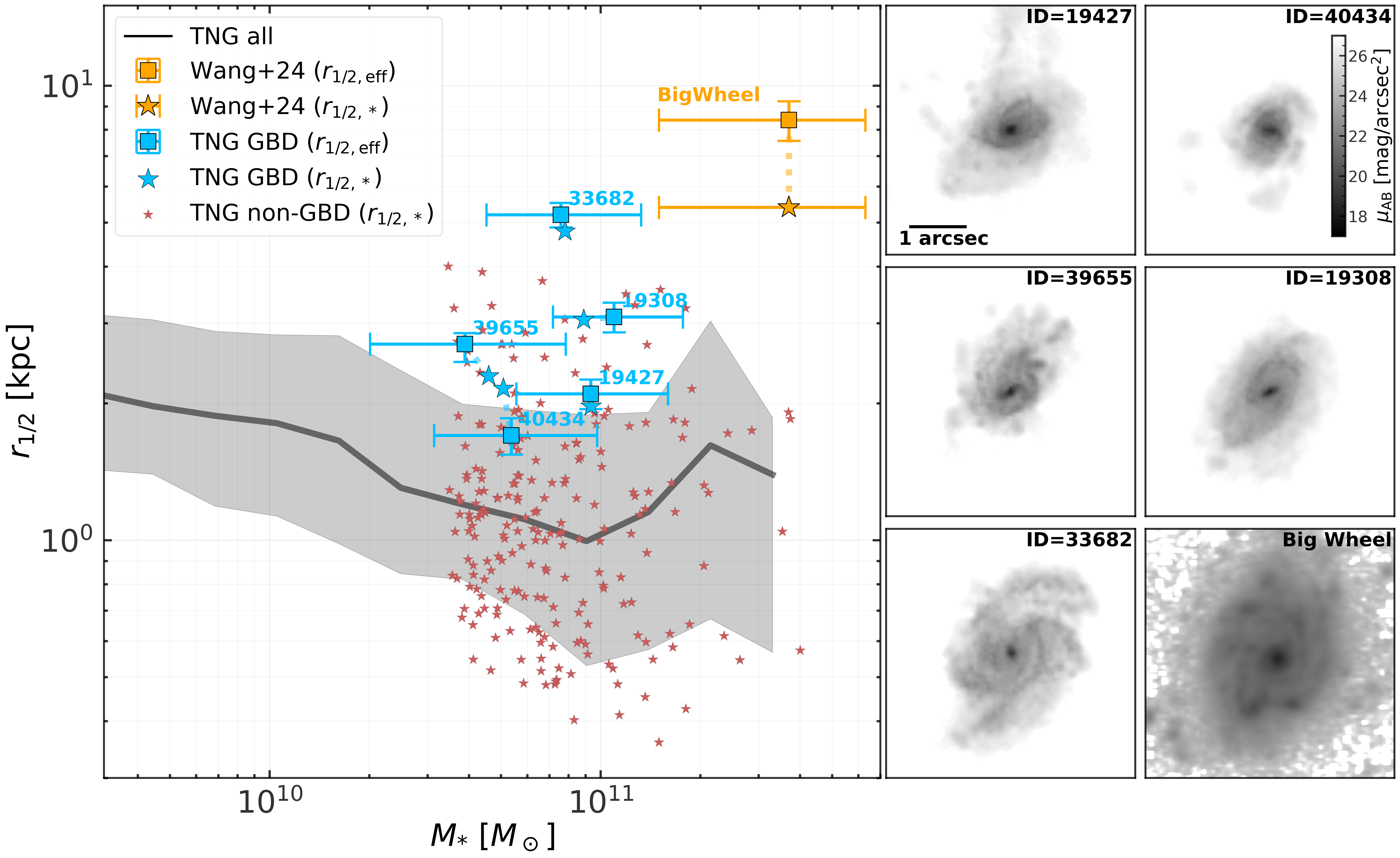}
    \caption{\textbf{Sizes - stellar mass relations at $z\approx3$ and mock images.} Our 5 candidates of {\it giant bulgeless disks} (GBDs) in the TNG100 simulation are shown with the blue symbols, and the observed Big Wheel (BW) galaxy is marked in orange. 
    Blue stars indicate the intrinsic half-stellar-mass radii and the stellar masses measured directly from the simulations. 
    Blue squares represent the apparent half-light radii based on mock JWST images in F322W2 band shown on the right side, and the stellar masses from fitting mock broad-band spectral energy distributions. 
    For the BW, the orange star stands for the half-mass radius estimate from \citep{Wang24}. 
    The simulated galaxies are adjusted to be of the same inclination angle and position angle as the BW.
    The small red stars represent normal disk galaxies in the same mass range as the GBDs but without the morphological constraints of extraordinarily large and bulgeless (non-GBDs), and the black line shows the median intrinsic size-mass relation of all the central galaxies in TNG100, with the shaded region representing the 16$^{\rm th}$ and the 84$^{\rm th}$ percentiles.
{\bf The largest GBD candidate (33682) is in the same ballpark as the BW, and all the candidates exhibit morphological and kinematical properties similar to the BW (summarized in \Tab{optmorph}).
Every galaxy has a dense inner component that appears to be a bulge and cannot be fully resolved in the JWST images, but photometric analysis reveals that the inner S\'ersic indices are all $\sim 1$, indicative of an inner disk.}
}
    \label{fig:size-mass}%
\end{figure*}

We identify GBD analogs in the TNG100 cosmological simulation \citep{Pillepich18b} at redshifts up to 3.
The selection criteria require a bulge-mass fraction below 5 percent \citep{Wang24}, a disk fraction greater than 80 percent (dominated by the thin disk), and half-stellar-mass radii exceeding the 84th percentile for galaxies in the same mass range of $\Ms > 10^{10.5}\Msun$.
These criteria capture the core features that their disks are well developed (extended and thin), and that they are limited in bulge component. 
The mass fractions of the morphological components are based on a new kinematic morphological decomposition algorithm that we developed \cite{Liang24}, as reviewed in the Methods section. 
These simulated GBDs are rare, accounting for only $\sim 1-2$ percent of all disk-dominated galaxies of similar mass within the simulation box. 
At $z=3$, only five candidates meet these criteria, forming the primary sample for this study.
As shown in \Fig{size-mass}, the simulated GBDs have sizes and stellar masses that are, while not as extreme, close to the observed BW galaxy. 
The largest example has a half-light radius of approximately $6\kpc$ and a stellar mass of $\sim10^{11}\Msun$, in the same ballpark as the BW. 
Here the size and mass estimates are derived from mock JWST observations, described in Methods. 
\Tab{optmorph} summarizes a comprehensive set of their morphological and kinematical parameters, highly similar to the observed one. 
In this study, we focus on the five GBD analogs at $z=3$ and incorporate lower-redshift samples at $z=2$ and $1$, as they provide qualitatively consistent interpretations.
The analysis is conducted under the assumption of a standard flat $\Lambda$-cold-dark-matter cosmology, as used in the TNG simulations, with virial mass defined by a spherical overdensity of 200 times the critical density.

\begin{table}[h!]
    \centering 
    \renewcommand{\arraystretch}{1.5} 
    \setlength{\tabcolsep}{1.25pt} 
    \small 
    \caption{{\bf Optical morphological properties of the observed Big Wheel (BW) galaxy and simulated GBDs at $z\sim3$ in the JWST F322W2 band.}}
    \label{tab:optmorph}
    \begin{tabular*}{\linewidth}{@{\extracolsep{\fill}}lccccccccc@{}}
        \toprule
        ID & $r_{1/2,\rm eff}$\footnotemark[1]  & $r_{\rm e,inner}$\footnotemark[2]  & $n_{\rm inner}$\footnotemark[3]  & $C$\footnotemark[4]  & $S$\footnotemark[5]  & $G$\footnotemark[6]  & $M_{20}$\footnotemark[7] & $v_{\rm rot}/\sigma$\footnotemark[8] & $v_{\rm rot}/\sigma$\footnotemark[9]  \\
         &  [kpc] &  [kpc] &  &  & & &  & (inner) & (total)\\
        \midrule
        BW & $8.4^{+0.8}_{-0.8}$  & $0.82^{+0.01}_{-0.01}$ & $0.72^{+0.03}_{-0.03}$ & $2.73^{+0.02}_{-0.02}$ & $0.04^{+0.01}_{-0.01}$ & $0.44^{+0.02}_{-0.02}$ & $-1.47^{+0.02}_{-0.02}$ & NA & $4.5^{+1.4}_{-1.1}$\\
        33682 & $5.2^{+0.3}_{-0.3}$  & $0.52^{+0.00}_{-0.00}$ & $0.69^{+0.00}_{-0.00}$ & $3.64^{+0.02}_{-0.02}$ & $0.11^{+0.01}_{-0.01}$ & $0.54^{+0.00}_{-0.00}$ & $-2.12^{+0.03}_{-0.03}$ & 2.1 & 4.1\\
        19308 & $3.1^{+0.2}_{-0.2}$ & $0.60^{+0.02}_{-0.02}$ & $1.05^{+0.08}_{-0.08}$ & $3.74^{+0.04}_{-0.04}$ & $0.03^{+0.00}_{-0.00}$ & $0.56^{+0.00}_{-0.00}$ & $-2.10^{+0.02}_{-0.02}$ & 3.1 & 3.9\\
        39655 & $2.7^{+0.1}_{-0.2}$  & $0.82^{+0.05}_{-0.05}$ & $1.88^{+0.02}_{-0.02}$ & $3.63^{+0.04}_{-0.04}$ & $0.04^{+0.01}_{-0.01}$ & $0.57^{+0.01}_{-0.01}$ & $-1.71^{+0.02}_{-0.02}$ & 2.0 & 7.7\\
        19427 & $2.1^{+0.1}_{-0.1}$  & $0.60^{+0.05}_{-0.05}$ & $0.73^{+0.35}_{-0.35}$ & $3.28^{+0.01}_{-0.01}$ & $0.07^{+0.00}_{-0.00}$ & $0.55^{+0.00}_{-0.00}$ & $-1.85^{+0.01}_{-0.01}$ & 3.1 & 6.8\\
        40434 & $1.7^{+0.1}_{-0.1}$  & $0.99^{+0.03}_{-0.03}$ & $0.58^{+0.02}_{-0.02}$ & $2.74^{+0.01}_{-0.01}$ & $0.01^{+0.03}_{-0.03}$ & $0.50^{+0.00}_{-0.00}$ & $-1.69^{+0.02}_{-0.02}$ & 2.2 & 4.8\\
        \bottomrule
    \end{tabular*}
\footnotetext[1]{Half-light radius.}
\footnotetext[2]{Effective radii of the inner S\'ersic components.}
\footnotetext[3]{S\'ersic index of the inner component (the embedded mini disk).}
\footnotetext[4,5,6]{Light concentration index ($C$), smoothness index ($S$), and Gini coefficient ($G$) \citep{Conselice03}.}
\footnotetext[7]{$M_{20}$ index for galaxy clumpiness \citep{Lotz04}.}
\footnotetext[8]{Ratio between rotational velocity and velocity dispersion $v_{\rm rot}/\sigma$ within the innermost 1kpc.}
\footnotetext[9]{Ratio between rotational velocity and velocity dispersion $v_{\rm rot}/\sigma$ for the entire simulated galaxy (considering particles within $5r_{1/2,\star}$. The ratio of BW is Measured by one dimensional H$\alpha$ kinematics profiles from observation \citep{Wang24}}
\end{table}

\section{Results}\label{sec: results}

\subsection{The environment for early giant disks}\label{sec:environment}

\begin{figure*}
    \centering
    \includegraphics[width=1\linewidth]{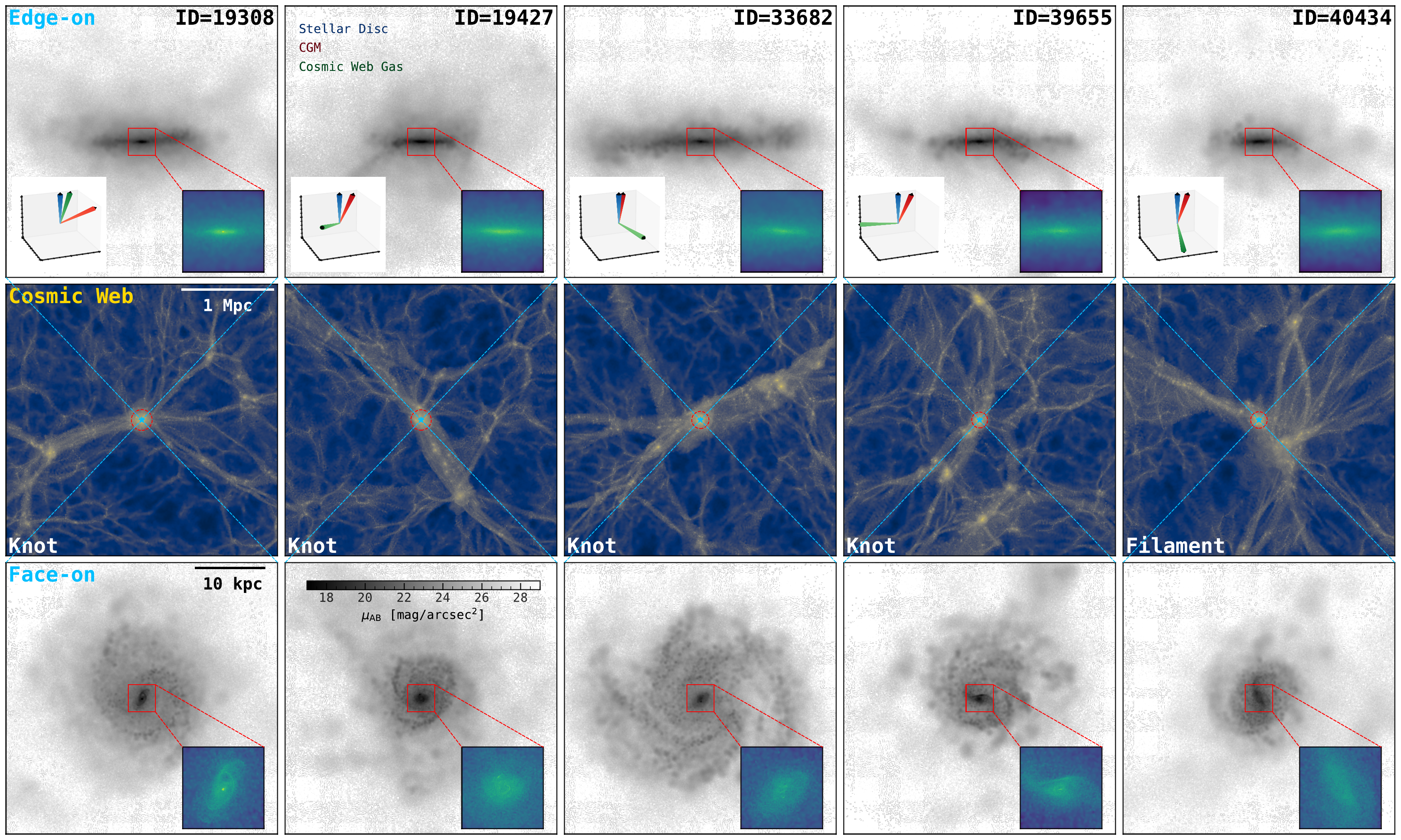}
    \caption{
\textbf{The large-scale environment and mock images of the simulated giant bulgeless disks (GBDs) at $z=3$.}
{\it top} and {\it bottom}: edge-on and face-on views of the galaxies.
The lower-left insets in the first row show the misalignments of the angular momenta of the galaxy, the CGM, and the gas in the cosmic web between 1 and 5 virial radii. 
Unlike the mock observations in \Fig{size-mass}, these synthetic images are generated with a finer spatial sampling than that of the JWST observation for revealing detailed structures.
The lower-right zoom-in insets show the baryonic surface density in the central region.
{\it Middle}: projected total mass distribution of the cosmic web around the GBD in a $3\Mpc$ cube. 
{\bf Every GBD has a compact mini disc in the center.
This universal feature may be marginally resolvable by JWST and may visually appear to be a bulge as shown in the mocks in \Fig{size-mass}.  
The CGM components in most cases are very well aligned with the disks, whereas the cosmic-web gas supplies are not necessarily kinematically coherent. 
All the GBDs reside in dense regions of the cosmic web that are either filaments or young cosmic knots that are still not fully virialized (i.e., proto-clusters).}
  }
    \label{fig:CosmicWeb}%
\end{figure*}

Being as massive as $\Mstar\sim10^{11}\Msun$ at $z=3$, these GBDs naturally reside in high-density peaks.
Their host halos, with masses of $\Mv\sim10^{12.5}\Msun$, correspond to density peak heights in the $\gtrsim2.5\sigma$ tails of the cosmic overdensity distribution at that epoch \citep{MBW10}.
Consequently, these giant disks are situated in the densest regions of the cosmic web.
Such dense environments are typically abundant in interactions and mergers, processes that are generally destructive to disk structures and conducive to bulge formation. 
This raises a critical question: what specific niche conditions allow these disks to grow so large while keeping their bulges minimal?

We perform a tidal-tensor classification of the cosmic web \citep{Liang24} and find that four of the five high-$z$ GBDs reside in cosmic knots, while one is located in a filament near a forming knot, as illustrated in \Fig{CosmicWeb}. 
Their large-scale neighbor densities are notably lower than those of normal disk galaxies of similar mass (referred to as non-GBDs, which are selected to be disk-dominated but without the other morphological constraints), as shown in \Fig{EnvirMergerCondition}. 
This indicates that early giant disks primarily exist in proto-galaxy clusters, which are gravitationally bound but not yet fully virialized. 
The observed BW follows this trend, located within a proto-cluster \citep{Wang24}.
The proto-cluster environment likely facilitates giant disk formation by ensuring a steady supply of gas along filaments, while destructive mergers have yet to occur.

\subsection{The assembly histories and halo conditions for early giant disks}\label{sec:halo}

\begin{figure*}[h!]
    \centering
    \includegraphics[width=1\linewidth]{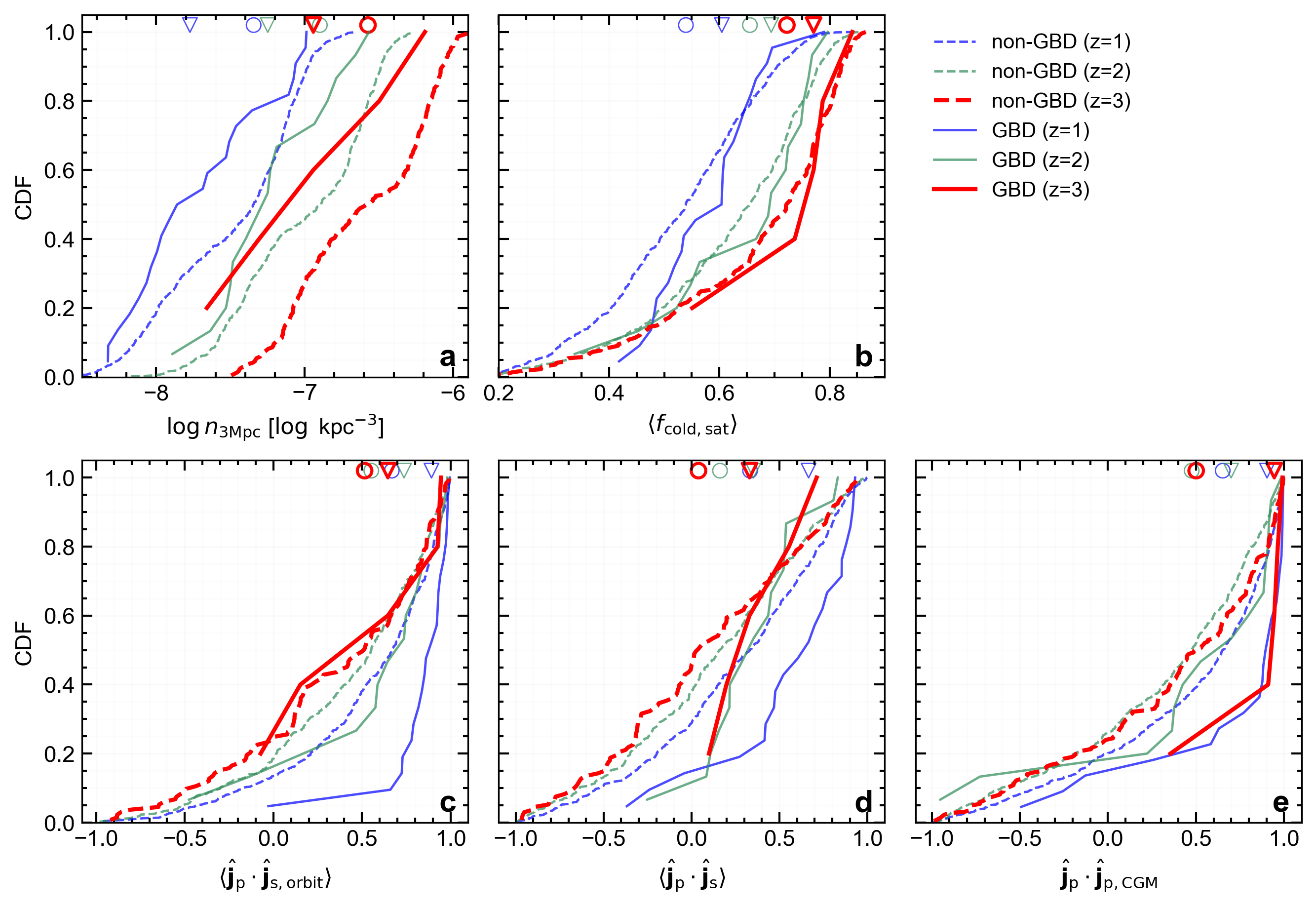}
    \caption{
\textbf{Cumulative distributions of environmental and merger properties for GBDs and non-GBDs at redshifts up to 3.} 
\textbf{a.} Halo number density within 3 Mpc, $n_{\rm 3Mpc}$, including all the neighbouring halos with masses exceeding 0.1\% of the host halo of interest. 
\textbf{b.} Average cold gas fraction of satellites that merged in the past, $\langle f_{\rm cold, sat} \rangle$. 
\textbf{c.} Average cosine of the angle between the angular-momentum vector of the cold baryon (stars + cold gas) of the primary progenitor, $\bm{j}_{\rm p}$, and the orbit angular-momentum vector of a merging satellite, $\bm{j}_{\rm s,orbit}$. 
\textbf{d.} Average cosine of the angle between the angular-momentum vector of the primary progenitor, $\bm{j}_{\rm p}$, and that of a satellite. 
The stellar-mass-weighted average cosine values are calculated for mergers within the last 4 dynamical times that penetrated within $10r_{1/2,\star}$ to their host center, with the angular-momentum vectors evaluated at their last peri-center passages.
\textbf{e.} Instantaneous cosine of the angle between the specific angular-momentum of the cold baryon, $\bm{j}_{\rm p}$, and that of the CGM, $\bm{j}_{\rm p, CGM}$, measured for all the gas between 0.1 and 1 $\Rv$.
The distributions at $z=1, 2$, and 3 are depicted in blue, green, red, respectively, with solid and dashed lines indicating the GBDs and non-GBDs, respectively, and open triangles and circles of corresponding colors at the top indicating the median values.
{\bf 
GBDs tend to reside in proto-clusters, are more likely to experience gas-rich, kinematically coherent mergers, and have remarkably coherent CGM.
}
}
    \label{fig:EnvirMergerCondition}
\end{figure*}

Further insights into the formation of GBDs can be obtained from the assembly histories and the structures of their hosting dark-matter and gaseous halos.
GBD progenitors are observed to accrete satellite galaxies that are notably richer in cold gas, as illustrated in \Fig{EnvirMergerCondition}b. 
Additionally, the orbital angular momenta of recent mergers exhibit a higher degree of alignment with the spin vector of the primary GBD progenitor compared to non-GBDs, as shown in \Fig{EnvirMergerCondition}c. 
 This alignment is quantified by the inner product of the unit angular-momentum vector of the GBD progenitor, $\hat{\bm{j}}_{\rm p}$, and that of the satellite orbit, $\hat{\bm{j}}_{\rm s,orbit}$, evaluated at the orbital pericenters. 
On average, $\hat{\bm{j}}_{\rm p} \cdot \hat{\bm{j}}_{\rm s,orbit}$ is approximately 0.6, 0.75, and 0.78 for GBDs at $z = 3$, 2, and 1, respectively, compared to 0.5, 0.55, and 0.65 for non-GBDs.
Hence, despite a redshift trend, at any given epoch, the GBD satellites deposit their cold-gas reservoir more coherently to the hosts than those of the non-GBDs. 
Furthermore, the internal baryonic spin vectors of the satellites also tend to align more closely with the primary GBD progenitors, as demonstrated in \Fig{EnvirMergerCondition}d.

Overall, the assembly histories of the GBDs are notably more quiescent: none of the GBDs experienced more than two major or minor mergers with stellar mass ratios exceeding $1:10$ within the last four dynamical times, whereas non-GBDs  experienced up to four such mergers. 
Additionally, the average stellar-mass ratio between the satellite and the GBD progenitor rarely exceeds $1:3$, whereas non-GBDs frequently undergo mergers with mass ratios approaching unity. 

Besides mergers, the spin vectors of the GBDs also exhibit strong alignment with those of the hot circum-galactic medium (CGM), as illustrated in the top panels of \Fig{CosmicWeb}. 
In contrast, the angular momentum of gas in the surrounding cosmic web is typically misaligned with both the disk and CGM spin vectors. 
This misalignment reflects the characteristic hot-mode accretion of these massive halos \cite{Dekel09}. 
Likely due to the presence of hot halos, which restrict the penetration of cold cosmic filaments that are not necessarily aligned, the internal halo environments of GBDs demonstrate remarkable coherence. 
\Fig{EnvirMergerCondition}e further quantifies the disk-CGM alignment: GBD hosts exhibit significantly more coherent CGM compared to non-GBDs, with a median $\hat{\bm{j}}_{\rm p} \cdot \hat{\bm{j}}_{\rm p,CGM}$ of 0.9, 0.6, 0.9 for GBDs at $z = 3$, 2, and 1, respectively, compared to 0.5, 0.5, and 0.65 for non-GBDs, where $\hat{\bm{j}}_{\rm p,CGM}$ is the unit angular-momentum vector of the surrounding gas at 0.1-1 virial radius. 
Gas cooled from the aligned CGM naturally contributes coherently to disk growth.
 
\begin{figure*}[h!]
    \centering
    \includegraphics[width=1\linewidth]{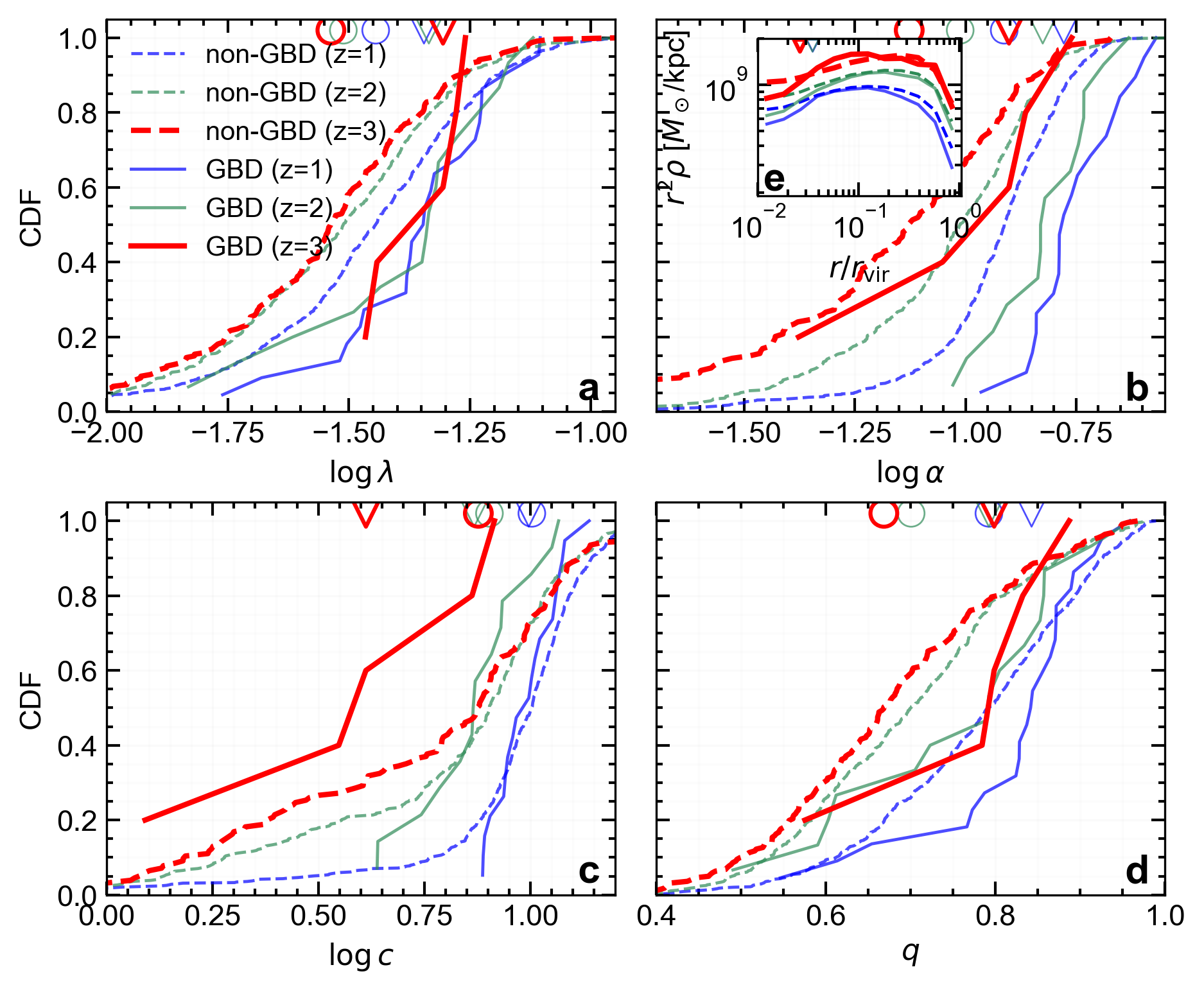}
    \caption{
\textbf{Cumulative distributions of the properties of the dark-matter halos hosting GBDs and non-GBDs}, with the same line and marker styles as \Fig{EnvirMergerCondition}: 
\textbf{a.} Spin parameter ($\lambda$). 
\textbf{b,c.} Shape index $\alpha$ and concentration parameter $c$ of the best-fit Einasto density profiles \citep{Einasto65}, with the inset \textbf{e} showing the corresponding average density profiles. 
\textbf{d.} 3D axis ratio ($q$, defined as intermediate axis : long axis).
{\bf  The dark-matter halos of GBDs exhibit higher spin, lower central densities, lower concentration, and more spherical shapes.}
    }
    \label{fig:HaloCondition}
\end{figure*}

Comprehensive measurements of the dark-matter halo structural properties (see Methods) reveal that the early giant disks reside in halos with distinctive internal structures and shapes.
First, the host halos of GBDs exhibit higher specific angular momentum. Their average spin parameter, $\lambda$, is approximately 0.2 dex higher than that of normal disk galaxies (\Fig{HaloCondition}a).
This aligns with the classical model of disk formation, in which galaxies inherit angular momentum from their host halos \citep{MMW98}.
However, unlike nearby disk galaxies, whose half-mass radii are typically $\sim 1\%$ of the virial radii of their halos \citep{Kravtsov13}, GBDs have sizes closer to $\sim 10\%$ of the halo virial radius. 
The relationship between halo spin and disk size remains debated, as cosmological simulations suggest that angular momentum is not conserved during galaxy formation \citep{Jiang19}. 
The TNG simulations, as used in this study, exhibit stronger correlations between disk size and halo spin compared to other models \citep{Yang23}.
Second, GBD hosts have significantly lower central dark-matter densities than non-GBDs, particularly obvious at $\sim 1\%$ of the virial radius as demonstrated in \Fig{HaloCondition}e , such that the overall Einasto profile index is $\sim$0.2 dex higher (\Fig{HaloCondition}b). 
Related, the dark-matter concentration parameters of GBD halos are generally lower than corresponding values for normal disks at $z=3$ (\Fig{HaloCondition}c).
Finally, the GBD halos are more spherical than those of the control sample, with the axis ratio $q\gtrsim 0.8$ (\Fig{HaloCondition}d).
These together indicate that the GBD host halos have shallower potential-well depth than the hosts of non-GBDs of similar mass, and that this shallower potential depth does not reflect recent accretion events which usually lead to non-spherical shapes. 
The increased sphericity is consistent with a more quiescent assembly history as discussed earlier.

\subsection{History and fate of the giant disks}\label{sec:evolution}

\begin{figure*}[h!]
    \centering
    \includegraphics[width=1\linewidth]{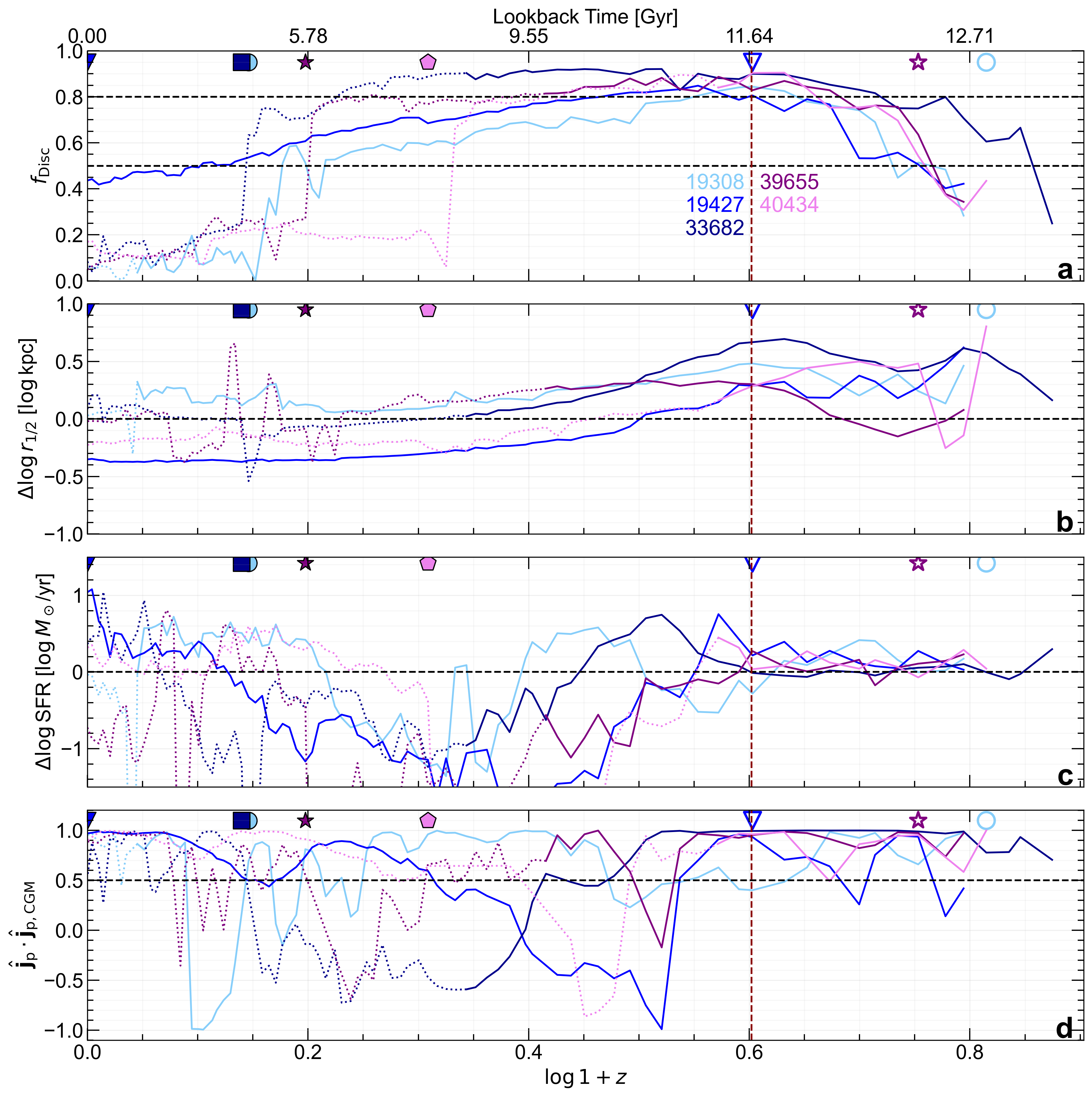}
    \caption{
    \textbf{Evolution of the simulated GBDs identified at $z=3$}.
     Each line represents one GBD, with solid and dotted segments indicating the phases where the galaxy is a central or a satellite, respectively. 
Open symbols mark the last major merger (if any) before $z=3$ (indicated by the vertical dashed line), while filled symbols denote the next major mergers. 
     \textbf{a.} disc mass fraction $f_{\rm Disc}$: horizontal dashed lines indicate $f_{\rm Disc}=0.5$ (threshold for disk dominance) and 0.8 (a selection criteria for GBDs). 
     \textbf{b.} excess half-stellar-mass size ($r_{1/2}$) relative to the median size-mass relation of the TNG100 simulation at the same redshift. 
     \textbf{c.} star-formation rate (SFR) excess relative to the median SFR-stellar mass relation (the star-forming main sequence) of the TNG100 simulation. 
     \textbf{d.} Cosine of the angle between the angular-momentum vector of the cold baryon and the CGM. 
The regime above the horizontal dash line indicate coherent CGM.
     {\bf The GBDs remain disk-dominated until the next destructive major merger, but their sizes become normal within 1-2 Gyr as disk instabilities develop and bulges grow. 
     During this time, the galaxies exhibit a tendency towards quenching, but star formation is rejuvenated by the subsequent gas-rich major merger (Panel c). 
    The CGM was coherent before the GBD phase (Panels a \& d), and became misaligned after $z\sim2$ when the disk sizes decrease to normal.}
     }
    \label{fig:history}%
\end{figure*}

As shown in \Fig{history}, the growth of the giant disks took place fairly rapidly.
At $z=3$, the GBDs lie on the star-forming main sequence with a specific star formation rate (sSFR) of $\sim 10^{-9}\rm{yr}^{-1}$, similar to that of the BW.
Their disk-dominated morphology remains intact until the next destructive major merger, but their exceptionally large sizes cannot be sustained. 
Over the course of $\sim 2\Gyr$ or a few dynamical times, as disk instabilities develop, these galaxies begin to converge toward the median size-mass relation.
As their half-mass radii shrink, the star formation rates also decline, eventually dropping approximately 1 dex below the main sequence. 
A subsequent gas-rich major merger rejuvenates star formation and simultaneously triggers the transition from disk-dominated to ellipsoidal morphology. By the time this process is complete, the GBD descendants have evolved into massive early-type galaxies, often serving as bright central galaxies in their environments.
Again, this verifies that the early giant pure disks are fundamentally different from nearby pure disks \citep{Kormendy10} and do not serve as progenitors of the local counterparts. 

The CGM became coherent early on, even before the GBD progenitor became disk dominated, and remained so throughout the phase when a GBD progenitor (or descendant) was exceptionally large and disk-dominated. However, the CGM became misaligned shortly after $z\sim2$, or when the disk size decreased to the ensemble average level, and the angle between the disk and CGM fluctuates significantly thereafter. 

\section{Discussion}\label{sec:discussion}

In this study, we provide a comprehensive analysis of giant bulgeless disks (GBDs) in the TNG100 cosmological simulation up to redshift 3. 
These early giant disks, like the observed Big Wheel (BW) galaxy, are outliers of the prevailing theoretical paradigm of disk formation, which posits that stable disks emerge following the formation of a compact stellar bulge. 
What sets them apart from normal high-$z$ disk galaxies that undergo the nugget-formation phase is their unique halo and environmental properties. 
They are almost exclusively located in actively forming cosmic knots, feature kinematically coherent circum-galactic medium (CGM), and experience mergers that are more gas-rich with better aligned orbital angular momentum and internal spin.
Additionally, their host dark-matter halos exhibit higher spin, lower halo concentration, lower central densities, and greater sphericity compared to their normal counterparts. 
Among these halo and environmental factors, which ones are the most important? 
We have analyzed the distributions of these factors in the parameter space spanned by size and disk fraction (in Supplement Information), and found that coherent CGM and mergers show the most clear diagonal trend. 
Dark-matter properties, especially lower concentration $c$ and higher spin $\lambda$, also play an important role in driving disks large, but do not necessarily prevent bulge formation. 
High cold-gas fraction in mergers and quiescent merger histories contribute to high disk fractions.

Are the GBDs stable? 
Their two-compoent Toomre $Q$ parameter \citep{Romeo11} remains nearly constant at $Q\sim 2-3$ within the disk range out to $\sim2 \rhalf$ (see Supplement Information). 
While this suggests a marginally stable state based on the classic instability criterion of $Q < 1$, previous simulation studies indicate that clumps can form precisely at $Q \sim 2-3$ \citep{Inoue16}. 
In contrast, the $Q$ values for non-GBDs exhibit clear monotonic increase with radius, slightly lower than those of GBDs within $\rhalf$, but higher in the outer disk up to $\sim2\rhalf$ where $Q$ reaches $\sim5$.
This implies that the high-$z$ GBDs are less stable than typical contemporary disks and are likely on the verge of clump formation.

We analyzed the non-parametric optical morphologies of the simulated GBDs in the JWST F332W2 band, summarized in \Tab{optmorph}. 
The $M_{20}$ indices, which quantify the contribution of the brightest regions to the overall galaxy morphology \citep{Lotz04}, have intermediate values around $-2$. 
These values indicate clumpy disk structures, falling between the lower values typical of ellipticals and higher values associated with mergers.
Similarly, the Gini coefficient ($G$), the smoothness parameter ($S$), and the light-concentration index ($C$) all lie in the intermediate ranges, with $\rm{G} \sim 0.4$-0.6,  $S\gtrsim 0$, and $C\sim2.5$-$4$, typical of disk galaxies \citep{Conselice03}. 
These characteristics closely resemble those of the observed BW.
However, as can be seen from \Fig{size-mass}, BW exhibits even greater clumpiness. 
This suggests that the level of clumpiness cannot be fully captured by current statistical metrics, and serves as a valuable benchmark for refining future numerical models.

It is intriguing that every GBD at $z=3$ in our sample features a dense disk embedded at its center.
This is particularly evident in the edge-on and face-on views shown in \Fig{CosmicWeb}, which are generated with higher resolution than JWST observations and without applying the JWST point-spread function (PSF), thereby preserving the simulation resolution. 
We further verified kinematically that these inner structures are indeed rotation supported with $v_{\rm rot}/\sigma\sim 2-3$ (\Tab{optmorph}), where $v_{\rm rot}$ is the rotational velocity and $\sigma$ is the velocity dispersion.
When convolved with the observational PSF and noise, these inner disks appear to be small bulges.
To further analyze their structure, we perform surface photometry on the mock observations using two S\'ersic components with {\tt GALFIT}. 
The best-fit S\'ersic index for the inner component is in most cases $\sim$0.6-1, with sub-kiloparsec effective radii. 
Remarkably, the observed BW exhibits a strikingly similar S\'ersic index of $\sim$ 0.7 for its inner component, along with a comparable size. 
This similarity suggests that inner disks may be a common feature in high-$z$ disk galaxies. 
While these disks remain barely resolved in JWST images, a S\'ersic index close to unity could serve as an indicator of their presence.
We caution though that a S\'ersic index of order unity is not a sufficient condition for disk morphology \citep{Gadotti09}.
The formation mechanisms and dynamical roles of these embedded disks lie beyond the scope of this study, but they present a compelling avenue for future explorations both theoretically and observationally.

\section{Methods}\label{sec：methods}

\subsection{Simulation} \label{sec:sim}

We use the publicly available IllustrisTNG simulations \citep{Marinacci18,Naiman18,Nelson18,Pillepich18a,Springel18}, a suite of magneto-hydrodynamic cosmological simulations with the moving-mesh code \texttt{AREPO} \citep{Springel10}. 
The simulations adopt the Planck 2015 \citep{Planck16} cosmology: $\Omega_{\mathrm{\Lambda}} = 0.6911$, $\Omega_{\mathrm{m}} = 0.3089$, $\Omega_{\mathrm{b}} = 0.0486$, $\sigma_{8} = 0.8159$, $n_{s} = 0.9667$, and $h = 0.6774$, and include comprehensive subgrid models for cooling, star formation, chemistry, and feedback from stars and black holes \citep{Weinberger17,Pillepich18b}. 
TNG galaxies are identified with the Friends-of-Friends (FoF) \citep{Davis85} and \texttt{Subfind} \citep{Springel01} algorithms and are linked across snapshots with the {\tt SUBLINK} merger tree algorithm \citep{Rodriguez-Gomez15}. 
For merger statistics, we adopt the {\tt Sublink\_gal} algorithm, a variant of {\tt Sublink} that builds the merger trees based on star particles and star-forming gas cells instead of dark-matter particles. 
TNG consists of three runs spanning a range of volume and resolution, TNG50, TNG100, and TNG300. 
This study requires decent numerical resolution for morphological analysis as well as large box volume to include sufficient number of the extremely large galaxies, so we adopted TNG100, which has a gas particle mass of $1.4\times 10^6\Msun$, a dark-matter particle mass of $7.5\times10^6\Msun$, and gravitational softening lengths of 0.740 comoving kpc for collisionless particles and 0.185 comoving kpc for gas particles. 

\subsection{Sample}\label{sec:sample}

We focus on well resolved galaxies in TNG100 with half-stellar-mass radii at least twice the collisionless softening length and containing more than 1000 stellar particles and 1000 DM particles within the virial radius. The following criteria are then applied to select the {\it giant bulgeless galaxies}:
(1) the candidate is a central with stellar mass $\Mstar > 10^{10.5}\Msun$, measured within $5\times r_{1/2}$.
(2) total disk mass fraction $f_{\rm Disc}\geq 0.8$, the thin disk fraction $f_{\rm ThinDisc}\geq 0.4$, and bulge fraction $f_{\rm Bulge}\leq 0.05$.
(3) half-stellar-mass radius $r_{1/2}$ greater than the 84th percentile for galaxies of the same stellar mass.
The measurements that used in these criteria are described in \se{decomposition}. 
Other systems satisfying (1) and $f_{\rm Disc}\geq 0.5$ but without other morphological constraints are labelled non-GBDs, which means that they are disk galaxies with similar stellar masses to GBDs but are normal looking in size and morphology. 
As such, we find 5, 11, and 22 GBDs and 221, 558, 782 non-GBDs at $z=3$, 2, and 1, respectively. 

\subsection{Morphological Decomposition}\label{sec:decomposition}
In this work, we use the public parameter-free package \texttt{MorphDecom} \citep{Liang24} that we developed to perform morphological decomposition. 
A key feature of \texttt{MorphDecom} is its ability to automatically determine the energy threshold for separating bulge from halo, and the circularity threshold for separating thin and thick disks, without aribitrary choices.  
For the energy threshold, \texttt{MorphDecom} follows the scheme in \cite{Zana22}, and for the circularity threshold, it uses the Gaussian-Mixture-Models algorithm (GMM). 
GMM has been shown to be an efficient method for morphological decomposition \citep{Du19}. 
However, previous works use this method by assigning Gaussian components to different morphological classes according to constant, user-specified energy and circularity thresholds, whereas \texttt{MorphDecom} improves this by automatically grouping the stars that do not belong to bulge or halo into thin and thick disks in the energy-angular-momenta space.  
 
\subsection{Synthetic Images}\label{sec:RT}
In order to emulate JWST observations at high redshift, we simulate the effect of dust on radiation as well as light distributions from stars and star-forming regions with the most updated version of \texttt{SKIRT} Monte Carlo radiative transfer code \citep{Camps20}. 
In particular, we adapt the pipeline of \cite{Costantin23} originally developed for TNG50 for our GBDs in TNG100, as follows.
(1) Old star particles with age $>10$ Myr are modeled as a stellar population with a high-resolution SED according to the initial mass functions of \cite{Bruzual03} and the \cite{Chabrier03}, while younger star particles will be modeled with an SED from the \texttt{MAPPINGS III} library \citep{Groves08}. For old stars, the inputs are their positions, current masses, metallicities, ages, and smoothing lengths, calculated using the 32$\pm$1 nearest stellar particles. 
For young stars, besides their positions, metallicities, and smoothing lengths, additional parameters for the \texttt{MAPPINGS III} library are required, including star formation rate, the compactness $C_0$ of the H II region (set to $\log_{10}C_0=5$), the pressure $P_0$ of the ISM (set to $\log_{10}[(P_0/k_{\rm B})/\text{cm}^{-3}\text{K}]=5$ where $k_{\rm B}$ is the Boltzmann constant), and the covering fraction of the photodissociation region $f_{\rm PDR}$ (fixed at 0.2).
(2) Dust is modeled using the properties of diffuse interstellar medium (ISM), since TNG lacks dust physics. 
We select cold and star-forming gas cells ($T<8000$K, SFR $>0M_\odot$/yr) as diffuse ISM. 
Next, we retrieve the positions, metallicities, gas mass density, and dust density as inputs. 
The dust density is determined by dust-to-metal ratio with given metallicity and mass density, i.e. $\rho_{\rm dust}=f_{\rm dust}Z\rho_{\rm gas}$. 
In this work, $f_{\rm dust}$ is fixed at 0.41. 
The dust composition model follows \cite{Zubko04}. The dust density distribution is discretized on an octree grid \citep{Saftly13} with a minimum refinement level of 3 and a maximum refinement level of 14. The maximum dust fraction in each cell is set to $10^{-6}$.
(3) For all the radiative-transfer simulations carried out for this work, we set $N_{\rm p}=10^8$ photon packets per galaxy. 
We use a logarithmic wavelength grid with 40 points, running from 0.08 to 10 $\mu$m for storing the mean radiation field, and a logarithmic wavelength grid with 200 points, running from 0.4 to 1000 $\mu$m for storing the mean dust emission field. 
For the instrument settings, we set the field of view 40 kpc with a pixel scale of 0.063 arcsec for \Fig{size-mass}, the same as the pixel scale of the Big Wheel observations. 
For the images in \Fig{CosmicWeb}, in order to reveal the fine structures of the galaxies, notably the inner mini disks, a finer pixel scale of 0.015 arcsec is used. 
A nested logarithmic grid is employed in instrument wavelength. 
The low-resolution part of this grid has 251 points, ranging from 0.32 to 4000 $\mu$m, whereas the higher resolution part, ranging from 0.32 to 12 $\mu$m, has 4001 wavelength points. 
We use all the NIRCam bands to construct images and mock SEDs.

\subsection{Dark-Matter Halo Properties}
(1) Virial radius $\Rv$ is defined as the radius of a spherical overdensity that is 200 times the comtemporary critical density of the universe. 
(2) Virial mass $\Mv$ is the total mass enclosed within $\Rv$. 
We use the fields of \texttt{Group\_R\_Crit200} and \texttt{Group\_M\_Crit200} from the public FoF halo catalog for $\Rv$ and $\Mv$, respectively.
(3) Halo concentration $c$ and shape index $\alpha$ characterize the dark-matter (DM) density profile, as in the Einasto parameterization \cite{Einasto65}, 
 \be
 \rho_{\rm DM}=\rho_{-2}\exp\left\{-\frac{2}{\alpha}\left[\left(\frac{r}{r_{-2}}\right)^\alpha-1\right]\right\}
 \ee
where $r_{-2}$ is scale radius at which the logarithmic densty slope is -2, concentration $c=\Rv/r_{-2}$, and $\rho_{-2}$ is the density at $r_{-2}$. 
In practice we fit the corresponding circular velocity profile $V_{\rm circ,DM}=\sqrt{GM_{\rm DM}(<r)/r}$ for the two parameters, as it yields less fitting error than fitting the density profile, where $M_{\rm DM}$ is the enclosed DM mass within radius $r$ calculated by integrating $\rho$. 
(4) Halo spin parameter is a dimensionless angular-momentum, defined as $\lambda=\jv/(\sqrt{2}\Rv\Vv)$ \citep{Bullock01}, where $\jv$ is the specific AM within the virial radius, and $\Vv$ is the circular velocity at the virial radius.
(5) The 3D axis ratios are constructed with the eigenvalues ($\lambda_1\geq\lambda_2\geq\lambda_3$) of the inertia tensor \citep{Allgood06}, 
$\mathcal{S}=(1/M)\sum_k m_k\boldsymbol{r}_{k,i}\boldsymbol{r}_{k,j}$, 
where the summation is over all the DM particles within the ellipsoid of interest, $\boldsymbol{r}_{k,i}$ is the component of the position vector of the $k$th particle along axis $i$, and $M=\sum_k m_k$ is the total mass within the ellipsoid. 
We measure the eigenvalues iteratively following the algorithm of \cite{Tomassetti16}.
The axis ratios are then calculated by $q=\sqrt{\lambda_2/\lambda_1}$, $p=\sqrt{\lambda_3/\lambda_2}$ and $s=\sqrt{\lambda_3/\lambda_1}$. 
(6) To classify the cosmic-web environment into void, sheet, filament, or knot, we use the T-web method \citep{Hahn07} based on the eigenvalues ($\lambda_i$) of the deformation tensor, 
$T_{ij} = \partial^2 \phi / \partial x_i \partial x_j$,
which is the Hessian of the gravitational potential $\phi$.
To determine the tidal tensor $T_{i,j}$ and the eigenvalues at each coordinate, we follow the steps of \cite{Martizzi19}, adapting the public code complementary to the work of \cite{Yang21}.
In this workflow, we adopt the following user-specific choices. 
First, we construct an overdensity grid with 512$^3$ bins using all the particles in the simulation.
The grid is smoothed with a Gaussian filter of scale $R_{\rm G}=0.5$ ckpc$h^{-1}$. 
The eigen-value threshold for environment classification is set to $\lambda_{\rm th}=0.4$, based on visual comparison of the resulting cosmic-web classification and the overdensity map of the simulation.
(7) We also quantify the large-scale environment of a galaxy using the local number density, by counting all nearby halos with mass greater than 0.1 percent of the halo of interest within 3 Mpc.

\subsection{Galaxy Properties}

(1) The stellar mass $\Mstar$ of a galaxy is measured by summing the masses of all stellar particles bound to the subhalo. 
(2) The mass fraction of morphological component $X$ is given by $f_{X} = M_{\star, X}/\Mstar$, where the stellar mass of $X$, $M_{\star, X}$, is measured using the decomposition method.
(3) The half-stellar-mass radius $r_{1/2,\star}$ is the radius containing half of the stellar mass of this halo. 
Both quantities are obtained using the \texttt{SubhaloMassType} and \texttt{SubhaloHalfmassRadType} fields from the public {\tt Subfind} halo catalog of the TNG100 simulation.
(4) The stellar masses from mock observations are obtained by fitting the mock SEDs following a similar procedure as used for the Big Wheel observation \cite{Wang24}, using the \texttt{prospector} code \citep{Leja17, Johnson21}. 
We assume the same signal-to-noise ratio as the data in \cite{Wang24}, keep the redshift fixed at $z=3$, and adopt a non-parametric SFH with continuity prior \citep{Wang23} and a binning strategy the same as that in \cite{Wang24}. 
We turn off the dust emission since only optical to near-infrared data is included, and exclude AGN since it is not implemented during radiative transfer.
Single stellar populations are modeled with \texttt{fsps} \citep{Conroy10}.
(5) The specific star formation rate (sSFR) is defind as $\text{sSFR}=\text{SFR}/\Mstar$ where $\text{SFR}$ is the star formation rate taken from the \texttt{SubhaloSFR} field from the {\tt Subfind} catalog, which is the summation of individual star formation rates of all the gas cells in a subhalo.
(6) The specific angular momentum (sAM) $\bm{j}$ of a galaxy is calculated by summing up the angular momenta of all the stars and cold gas within $r_{1/2,\star}$ and normalizing by the total cold baryonic mass, i.e. $\bm{j}=\sum_i m_i\bm{r}_i \times \bm{v}_i / \sum_i m_i$, where $\bm{r}_i$ and $\bm{v}_i$ are the position and velocity of the baryon particle $i$.  
In \Fig{EnvirMergerCondition}c, we denote the sAM for the primary progenitor of a galaxy $\bm{j}_{\rm p}$ and that of its satellite $\bm{j}_s$.
(7) The orbit angular momentum of a satellite is defined as $\bm{j}_{\rm s, orbit}=\bm{r}_{\rm s}\times \bm{V}_{\rm s}$ where $\bm{r}_{\rm s}$ and $\bm{V}_{\rm s}$ are the satellite’s position and velocity with respect to its central, obtained from the \texttt{SubhaloPos} and \texttt{SubhaloVel} fields in the {\tt Subfind} catalog, defined as the position of the particle with minimum gravitational potential energy in the halo, and the mass-weighted sum of all particle velocities, respectively.
(8) The stellar age is defined as the difference between the cosmic age at the time of interest and that when a star was formed, using the \texttt{GFM\_StellarFormationTime} field from the \texttt{PartType4} (stars) snapshot data. We exclude wind particles in the stellar-age analysis.
(9) The radial and rotational velocities are obtained by decomposing the velocity vector $\bm{v}$ into $(v_{\rm R}, v_{\rm rot}, v_{\rm z})$, where $v_{\rm R}$, $v_{\rm rot}$, and $v_{\rm z}$ represent the radial, rotational, and azimuthal velocities, respectively. Here, the velocity $\bm{v}$ is measured after rotating the velocity vectors such that $v_{\rm z}$ is parallel to the AM of stellar particles within $5r_{1/2,\star}$. The radial velocity dispersion $\sigma$ for each particle is calculated as 
$\sigma = \sqrt{(1/N) \sum (v_{\rm R} - \langle v_{\rm R} \rangle)^2}$
 using the nearest 16 neighbors. The velocity dispersion profile or velocity dispersion for the galaxy is therefore calculated by taking the mass-weighted average values of velocity dispersion for the particles of interests.
(10) The CGM gas is defined as all gas within $0.1 - 1 R_{\rm vir}$ for central galaxies. For their progenitors in the satellite stage (which is rare), CGM gas is defined as all gas beyond $5r_{1/2,\star}$.
(11) The Toomre Q parameter, $Q=\sigma\kappa / (\pi G\Sigma)$, describes the local instability of a disk \citep{Toomre64}, where $\sigma$ is the radial velocity dispersion, $\Sigma$ is the surface density, 
$G$ is the gravitational constant, and $\kappa$ is  the epicyclic frequency,
$\kappa =\sqrt{2} \left[\Vc^2 / R^2+(\Vc/R)(\rmd \Vc/\rmd R)\right]^{1/2}$, 
with $\Vc(R)$ the mid-plane circular velocity of the whole system, evaluated using the method for the circular velocity in the morphological analysis \citep{Liang24}.  
For estimating the effective $Q$ in two-component disks of both stars and gas, we follow the Wang-Silk approximation \citep{Wang94,Romeo11}, 
\be
\frac{1}{Q_{\text{comp}}} = 
\begin{cases} 
\frac{W}{Q_\star} + \frac{1}{Q_{\rm gas}}, & (Q_\star > Q_{\rm gas}), \\ 
\frac{1}{Q_\star} + \frac{W}{Q_{\rm gas}}, & (Q_\star < Q_{\rm gas}), 
\end{cases}
\ee
where $W = (2 \sigma_\star \sigma_{\rm gas})/(\sigma_\star^2 + \sigma_{\rm gas}^2)$.
Disk instabilities are expected to develop when $Q<Q_{\rm crit}\simeq1$. 
The critical value may be larger than unity for realistic systems \citep{BT08,Fujii11,Michikoshi14,Hu16}. 
For two-component disks, $Q_{\rm crit}$ should be raised to 2-3 \citep{Elmegreen11,Inoue16}.

\subsection{Merger statistics}\label{sec: merger}
We trace the progenitors and their secondary companions using the {\tt SUBLINK\_GAL} merger tree. 
(1) The {\it merger time} is defined as the time when the satellite galaxy can no longer be traced. 
(2) According to the {\it mass ratio}, $R_\star$, between the stellar mass of the primary progenitor and the peak stellar mass ever reached by the secondary, we call the mergers major ($R_\star>1/4$) or minor ($1/10<R_\star<1/10$). 
We exclude spurious mergers with either less than 50 stellar particles, $R_\star>1$, or subhalos of non-cosmological origin as marked by the \texttt{SubhaloFlag} field in the {\tt Subfind} catalog. 
Only mergers occurring within the last four halo dynamical times are considered ($\tdyn=(3\pi/16 G\rhobar)^{1/2}$, with $\rhobar$ 200 times the critical density $\rho_{\rm crit}(z)$). 
The mean merger mass ratio is then calculated as the descendent-stellar-mass-weighted $R_\star$ over all the mergers in the time window that a galaxy underwent.
(3) The {\it merger number} is the number of major and minor mergers that occur within four dynamical times.
(4) The {\it mean cold gas fraction} of satellites $\langle f_{\rm cold, sat} \rangle$ is based on the \texttt{MeanGasFraction} dataset in the public TNG100 merger-history catalog, where cold refers to a number density $n_{\rm H}>0.13$ cm$^{-1}$. 
This fraction is calculated as a mass-weighted average of all the satellites that have merged with the galaxy, using the peak stellar mass as the weight. 
(5) We calculate the {\it orbit misalignment} and the {\it spin misalignment} using the sAM vectors defined earlier, as the inner products $\hat{\bm{j}}_{\rm p}\cdot \hat{\bm{j}}_{\rm s, orbit}$ and $\hat{\bm{j}}_{\rm p}\cdot \hat{\bm{j}}_{\rm s}$, respectively. 
We calculate these misalignments at the snapshot when the satellite reaches its pericenter if the pericenter is within $10 r_{1/2,\star}$; otherwise, we use the last snapshot before merger. 
If both events occur beyond $10r_{1/2,*}$, we exclude this subhalo from this analysis.

\subsection{Optical Morphologies} \label{sec:optmorph}
We measure the optical morphologies from mock observations.
To this end, we first use {\tt SKIRT} to generate the mock images, following the observation strategy of the Big Wheel galaxy as in the JWST GO 1835 program (PI: Cantalupo), which has 1632 seconds of exposure for two NIRCam filters, F150W2 and F322W2.
Then, we convolve the SKIRT images with the empirical JWST NIRCam Point-Spread Function (PSF) measured by stacking the observed stars. 
We apply Poisson noise of the source and background noise based on the real observation image cutoff of the Big Wheel excluding sources and artifacts.

For non-parametric morphological measurements, we use the \texttt{statmorph} package \citep{Rodriguez-Gomez19}. 
Prior to the calculations, we first use \texttt{photutils} \citep{Bradley24} for source extraction and segmentation, using the standard deviation of our background noise as the threshold for source detection on the image that has been smoothed by a Gaussian kernel. 
We then put our JWST mocks to \texttt{statmorph} with the segmentation image.
We defined the effective radius $r_{1/2, \rm {eff}}$ as the semi-major axis of the ellipse that encloses half of the total light of the galaxy. 
We find that the PSF smoothing effect can lead to an overestimation of 14\% for exponential disks with an intrinsic half-light size of 0.3' under the observational condition, and apply the correction accordingly. 
For the uncertainty in size, we start with a 10\% relative fluctuation to account for systematics introduced by non-parametric ways to determine galaxy boundary, and propagate the error with 500 random noise realizations.
The other non-parametric morphological properties include concentration $C$, smoothness $S$, Gini coefficient, and the $M_{20}$ index. 
The concentration index $C$ is the ratio of the 80\% to 20\% curve-of-growth radii \citep{Conselice00}, with a larger value indicating a more concentrated light distribution. 
The smootheness or clumpiness index $S$ is computed by comparing the image to a smoothed, low-resolution image \citep{Conselice03}, with larger values referring to more clumpy structures.
The Gini coefficient quantifies the inequality in pixel flux by weighting the data according to its rank \citep{Lotz04}, ranging from zero, indicating that all pixels have the same flux, to unity, reflecting maximal flux disparity. 
The $M_{20}$ statistic is the normalised second-order moment of the brightest 20\% of the total flux, providing an alternative measure of concentration that does not assume the peak flux is centered \citep{Lotz04, Wang24PSF}. 
A high $M_{20}$ typically indicates more extended bright regions, as seen in mergers or disturbed morphologies.

We use \texttt{GALFIT} \cite{Peng02,Peng10} to perform two-component decomposition on the JWST F322W2 mocks of TNG100 galaxies.
We use double S{\'e}rsic fit to model the light profile of galaxies. 
The S{\'e}rsic model \citep{Sersic63} is defined as
\be
I(r) = I_{\rm e} \exp\left\{-b_n\left[\left(\frac{r}{r_{\rm e}}\right)^{1/n}-1\right]\right\}
\ee
where $I_e$ is the intensity at the effective radius $r_e$, $n$ is the S{\'e}rsic index, and $b_n$ is a constant that depends on $n$. 
We do not constrain the S{\'e}rsic index $n$ of the inner component. 
The position angle, the axis ratio, and the flux of the two components are allowed to vary independently.

\section{Supplementary Information}\label{sec: supp}
Here, we further compare the properties of giant bulgeless disks (GBDs) and those of normal disk galaxies in the same mass range (non-GBDs). 
First, we examine the radial profiles, including age specific star formation rate, gas-phase metallicity\footnote{The gas-phase metallicity is defined as log(O/H)+12, measured as the number ratio between oxygen and hydrogen atoms in logarithmic scale. The mass fraction of oxygen and hydrogen in each gas cell are obtained using \texttt{GFM\_Metals}. The number fraction is then converted by the nucleus number of H and O, 1 and 16.}, and the two-component Toomre $Q$ parameter. 
Second, to identify the most important halo and environmental factors that drive GBD formation, we analyze the distributions of these quantities in the spaces spanned by morphological parameters. In particular, we consider size, disk mass fraction $f_{\rm disk}$, and bulge mass fraction $f_{\rm bulge}$.

\begin{figure*}[h!]
    \centering
    \includegraphics[width=0.9\linewidth]{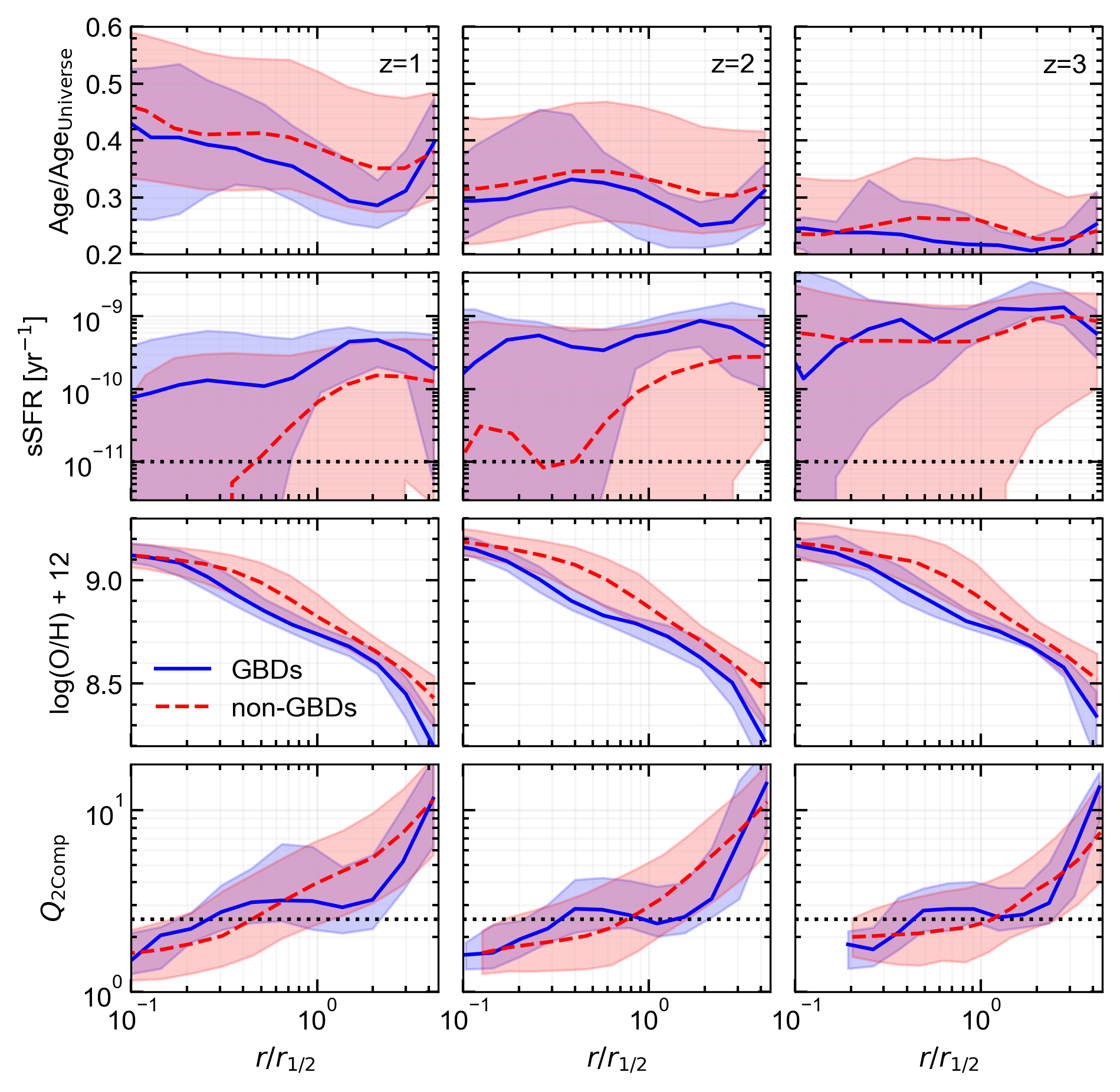}
    \caption{\textbf{Radial profiles for GBDs and normal disk galaxies of similar mass (non-GBDs) at redshifts up to 3.}
    \emph{First row:} Stellar age normalized by the universe age at each redshift. 
    \emph{Second row:} Specific star formation rate, sSFR, obtained by star formation rate of gas particles at each bin and stellar mass in the same bin. The horizontal black dotted lines represent the quenching threshold of sSFR=$10^{-11}$yr$^{-1}$.
    \emph{Third row:} Gas phase metallicity, log(O/H)+12.
    \emph{Fourth row:} Two-component Toomre stability index, $Q_{\rm 2 Comp}$. The horizontal black dotted lines indicate $Q_{\rm crit}=2.5$, an approximate stability threshold according to various simulation studies (see references in the paper). 
The median profiles of GBDs and non-GBDs are represented by blue and red lines, respectively, with the shaded areas indicating 16th and 84th percentiles. 
\textbf{The GBDs are generally younger, with higher star-formation rate, and lower metallicity compared to non-GBDs, and are marginally stable. They seem to show two relatively young components at $\sim 0.3\rhalf$ and $\sim 2\rhalf$.
}  
    }
    \label{fig:profiles}
\end{figure*}

\subsection{Radial profiles}
In \Fig{profiles}, we present the radial profiles of GBDs and non-GBDs at $z=1,2,3$. 
Overall, the GBDs exhibit younger stellar populations, higher sSFR, and lower gas-phase metallicities. 
Closer examination reveals that the GBDs also show distinct shapes in the profiles -- they have two components separated at $r \sim 0.6-0.9\,r_{1/2}$. 
Notably, their age profiles exhibit a young stellar population in the outer region at $\sim2 r_{1/2}$.
There is an older stellar population around $0.3-0.6\rhalf$, whereas the innermost part within $\sim 0.3 \rhalf$ is relatively young again. 
This ``$N$-shaped" behavior (from inner part to the outskirt) is particularly obvious at $z=2$, and qualitatively holds at other epochs as well. 
The sSFR profiles of GBDs show consistent behaviors, with double peaks in both inner and outer parts, at  $\sim 0.3 \rhalf$ and $\sim2 \rhalf$, respectively. 
In contrast, non-GBDs exhibit positive radial sSFR gradients, with suppressed star formation in their centers. 
The inner sSFR peak corresponds nicely to the inner disks, which we find to be a universal feature in high-$z$ GBDs, as illustrated in the paper. 
The gas-phase metallicity profile of non-GBDs follows a broken power law with a characteristic scale at $r \sim 0.4 r_{1/2}$, whereas GBDs seem to show two bumps in the profile, at $r \sim 0.2\rhalf$ and $\sim 2\rhalf$, respectively. 
All these features collectively reveal that GBDs host relatively young inner disks, besides an extended outer disk -- a unique feature that is worth theoretical and observational follow-ups.

As we discussed in the paper, the $Q_{\rm 2Comp}$ profiles of GBDs are quite flat at $\sim 2-3$ within $\sim2\rhalf$.
This manifests a marginally stable status that can lead to clump formation later on, consistent with the fact that $\fdisk$ decreases towards lower redshifts before any major merger occurs. 
GBDs are less stable in the outer disks at $r\sim2\rhalf$ compared to non-GBDs, the latter of which show monotonically increasing $Q_{\rm 2Comp}$ with radius.

\subsection{Feature importance}
\begin{figure*}[h!]
    \centering
    \includegraphics[width=1\linewidth]{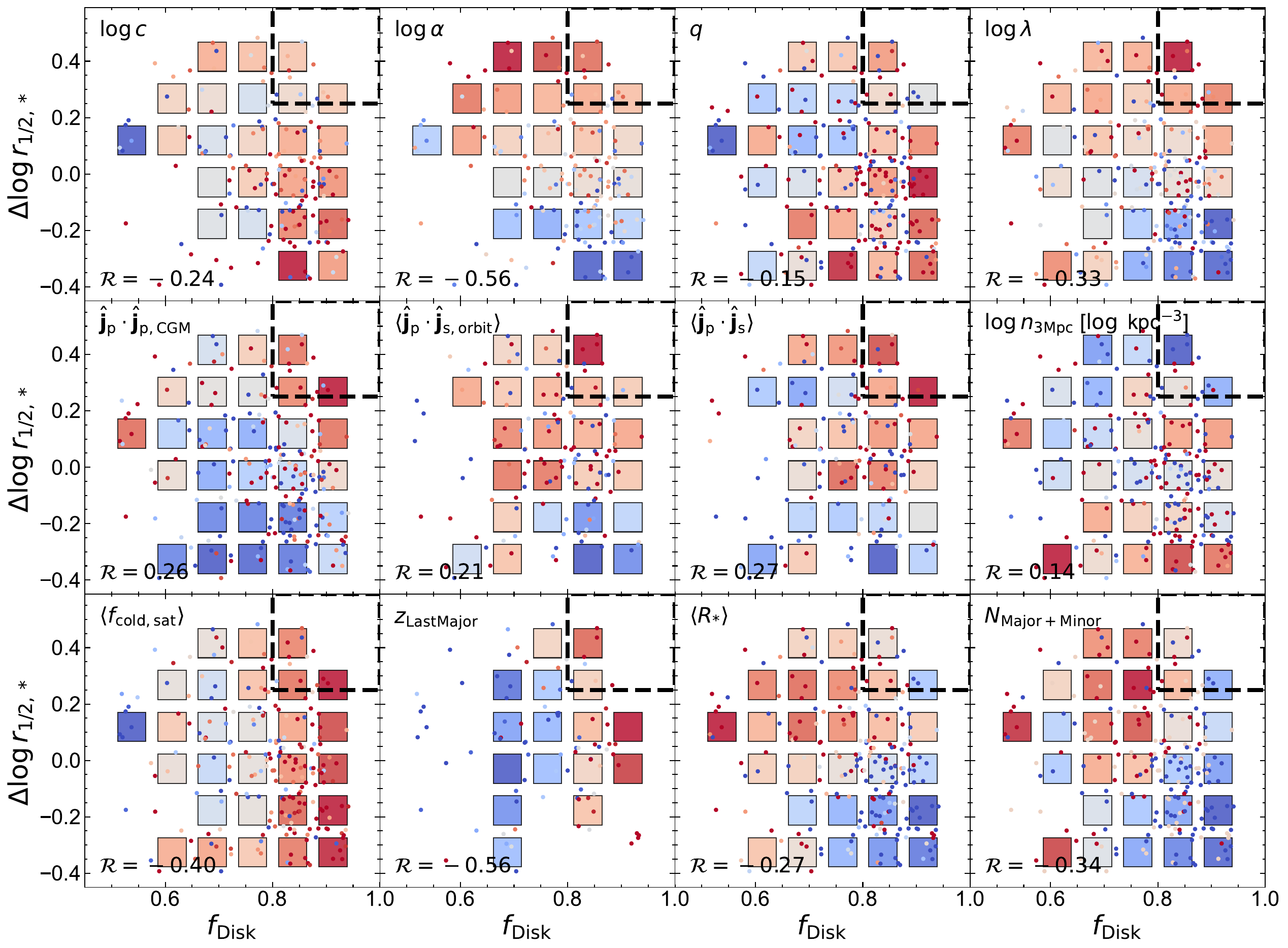}
    \caption{
\textbf{Diagnostics of the importance of relevant actors for GBD formation}, in terms of their distributions in the $\Delta \log r_{\rm 1/2}$ - $f_{\rm disk}$ plane, at $z=3$ in the TNG100 simulation, including {\it
dark-matter-halo properties} (concentration $c$, Einasto-profile shape$\alpha$, axis ratio $q$, spin $\lambda$), 
the {\it metrics for alignment} between the galaxy and hosting CGM or merging satellites (instantaneous cosine of the angle between the galaxy's angular-momentum vector and the angular-momentum of the hosting CGM $\hat{\bm{j}}_{\rm p} \cdot \hat{\bm{j}}_{\rm p,CGM}$, the average cosine of the angle between the angular-momentum vector of the galaxy and the orbital angular-momentum of a merging satellite $\langle \hat{\bm{j}}_{\rm p} \cdot \hat{\bm{j}}_{\rm s,orbit} \rangle$, and
the average cosine of the angle between the angular-momentum vector of the galaxy and that of a merging satellite $\langle \hat{\bm{j}}_{\rm p} \cdot \hat{\bm{j}}_{\rm s} \rangle$, where average cosine values are stellar-mass weighted, for all the mergers during the last 4 dynamical times that penetrated within 10$\rhalf$), 
the {\it large-scale number density} (number density within 3 Mpc, $n_{\rm 3Mpc}$, of all the neighbouring halos with masses exceeding 0.1\% of the halo of interest), 
and 
{\it merger statistics}
(average cold gas fraction of all the mergers in the past, $\langle f_{\rm cold,sat} \rangle$, 
redshift of last major merger $z_{\rm LastMajor}$, average stellar mass ratio of all the mergers $\langle R_{*} \rangle$, and the number of major and minor mergers $N_{\rm Major+Minor}$ during the last 4 dynamical times).
The color of a square represents the mean value of galaxies within a box of 0.11 dex in $f_{\rm disk}$ and 0.22 dex in $\Delta \log r_{1/2}$ around the center position of each square, calculated from at least 10 samples -- bluer means lower values and redder means stronger values. These mean values are obtained using locally weighted regression smoothing in Python package \texttt{LOESS}. 
The colors of the individual data points show the unsmoothed values, using the same color scale as the squares. 
The multiple correlation coefficients are displayed in the lower-left corners. 
The black dashed rectangle highlights the regime that corresponds roughly to GBDs.
}  
    \label{fig:rh_fdisk_X_z3}
\end{figure*}

\begin{figure*}[h!]
    \centering
    \includegraphics[width=1\linewidth]{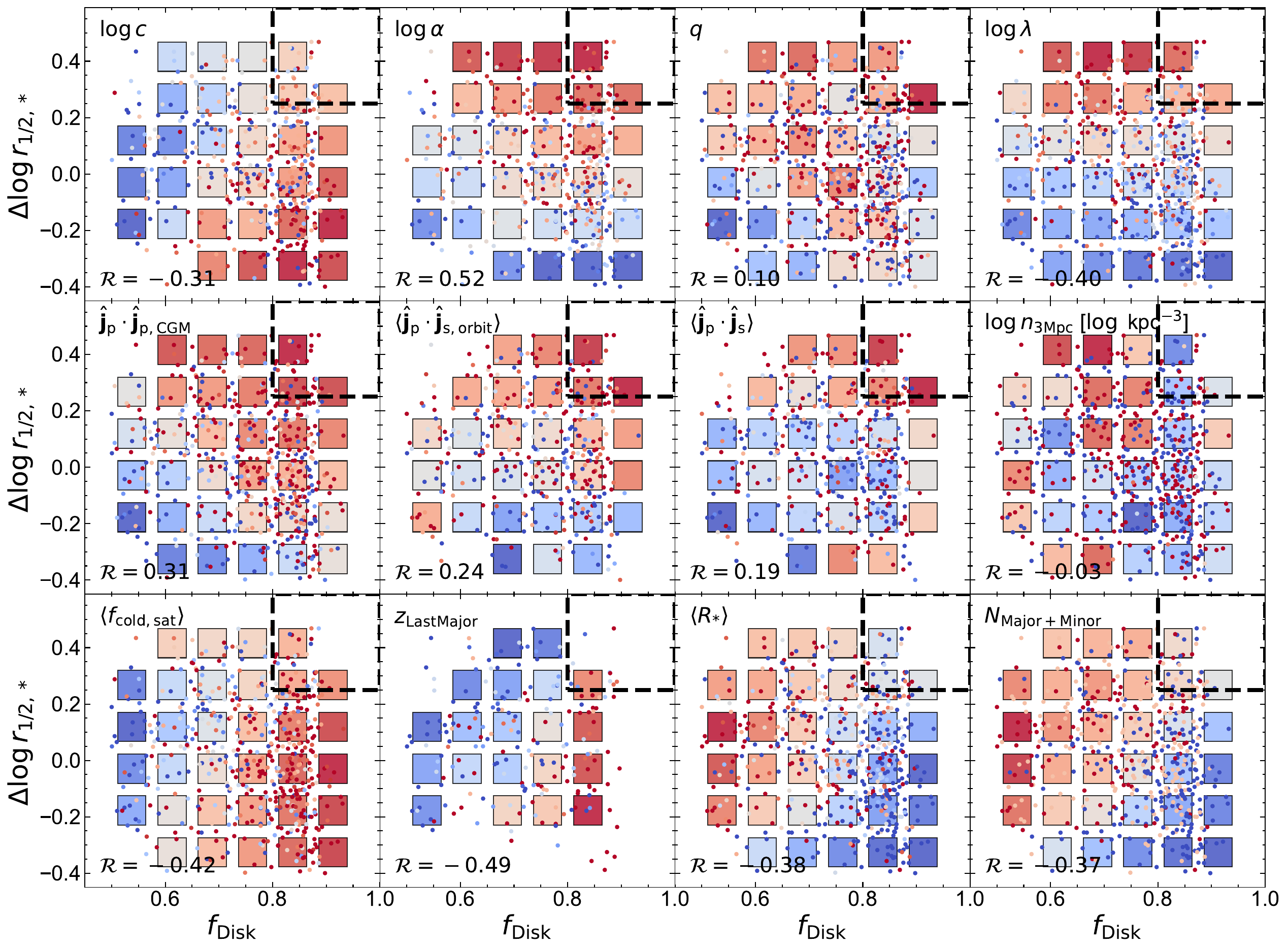}
    \caption{Same as \Fig{rh_fdisk_X_z3}, but for $z=2$, and the average is taken over 10 galaxies  (instead of 5) for each square.}  
    \label{fig:rh_fdisk_X_z2}
\end{figure*}

\begin{figure*}[h!]
    \centering
    \includegraphics[width=1\linewidth]{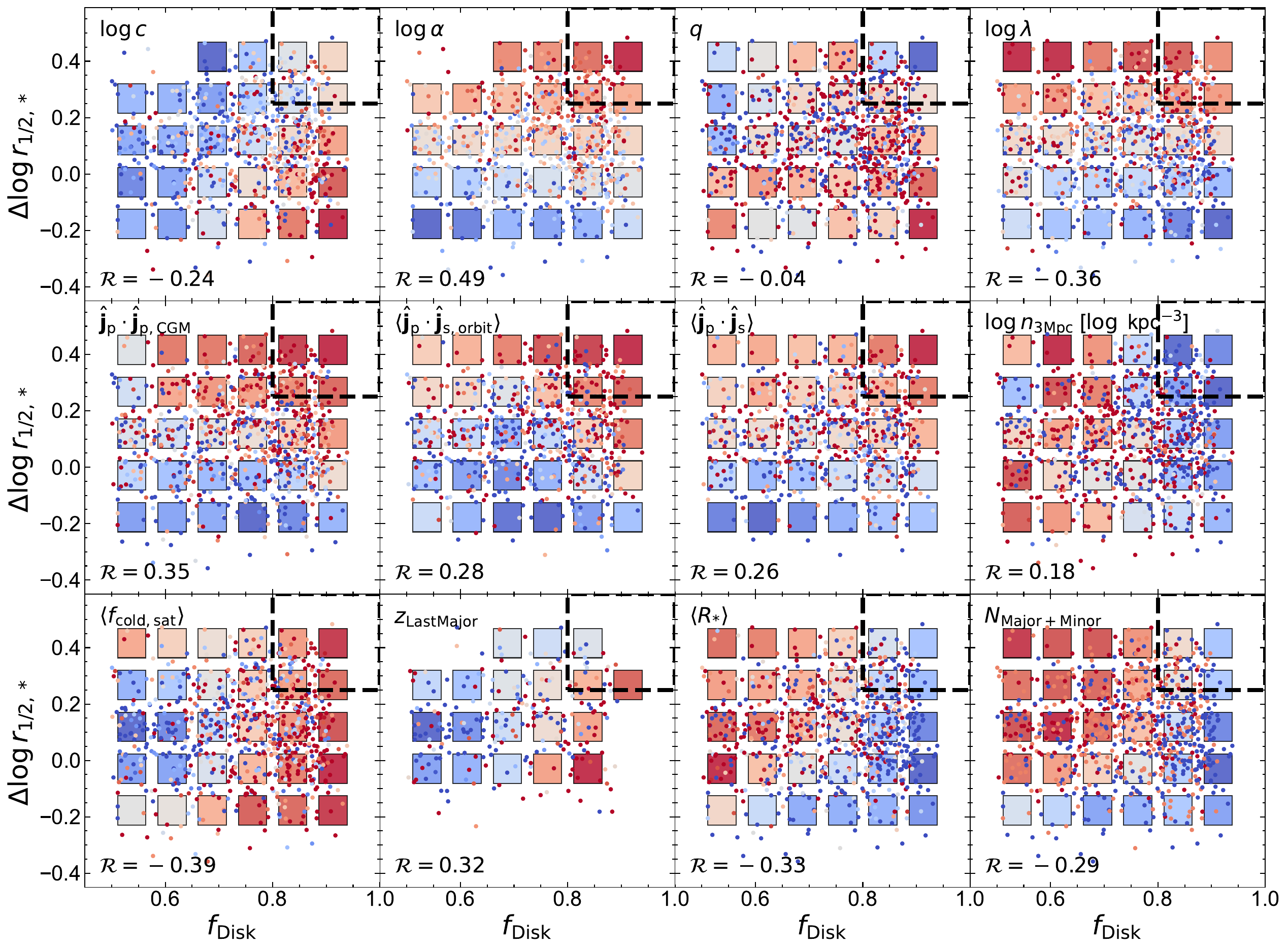}
    \caption{Same as \Fig{rh_fdisk_X_z2} but for $z=1$.}  
    \label{fig:rh_fdisk_X_z1}
\end{figure*}

\begin{figure*}[h!]
    \centering
    \includegraphics[width=1\linewidth]{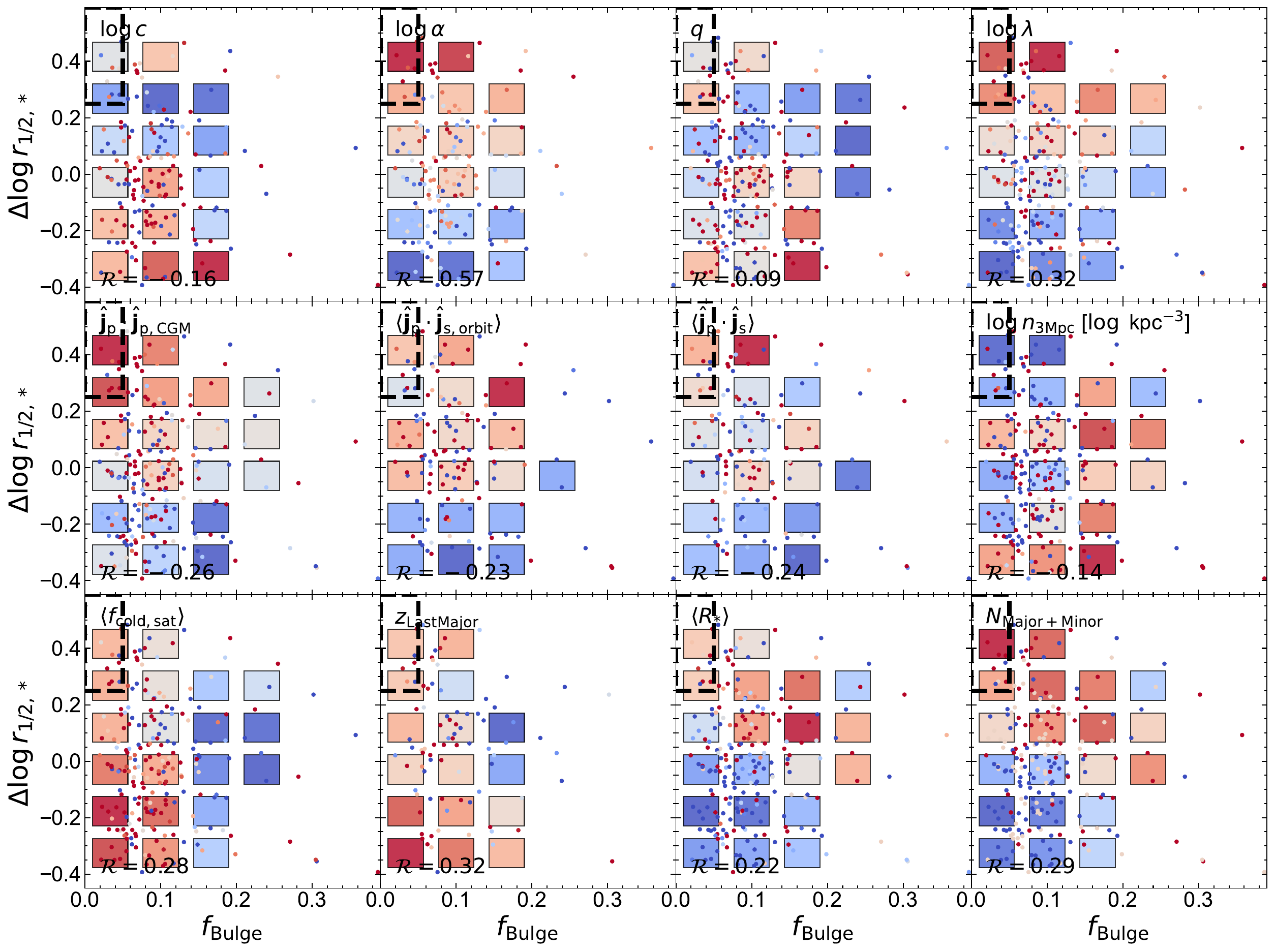}
    \caption{Similar to \Fig{rh_fdisk_X_z3}, but for the $\Delta \log r_{\rm 1/2}$ - $f_{\rm bulge}$ space.}
    \label{fig:rh_fbulge_X_z3}
\end{figure*}

\begin{figure*}[h!]
    \centering
    \includegraphics[width=1\linewidth]{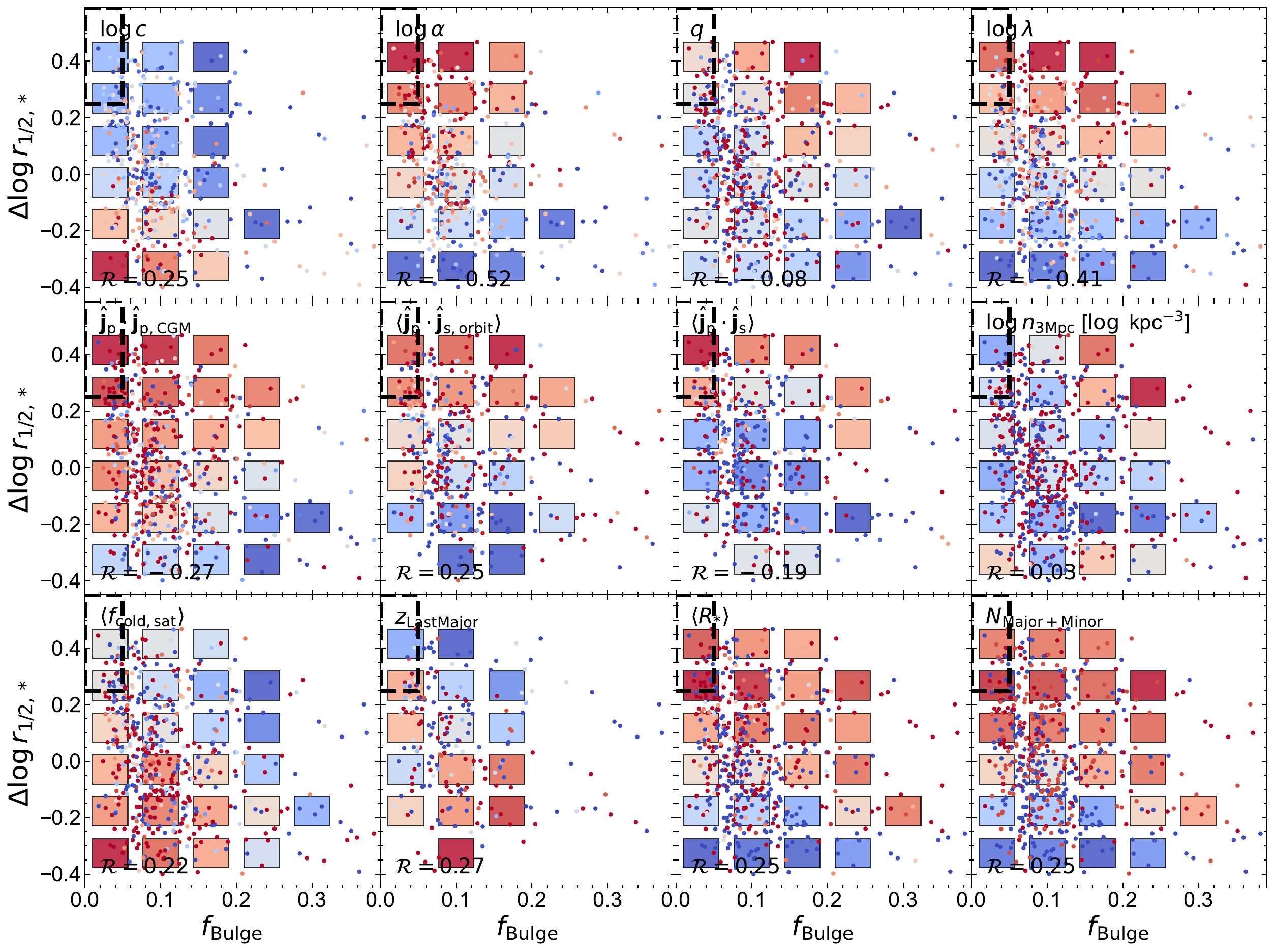}
    \caption{Similar to \Fig{rh_fdisk_X_z2}, but for the $\Delta \log r_{\rm 1/2}$ - $f_{\rm bulge}$ space.}  
    \label{fig:rh_fbulge_X_z2}
\end{figure*}

\begin{figure*}[h!]
    \centering
    \includegraphics[width=1\linewidth]{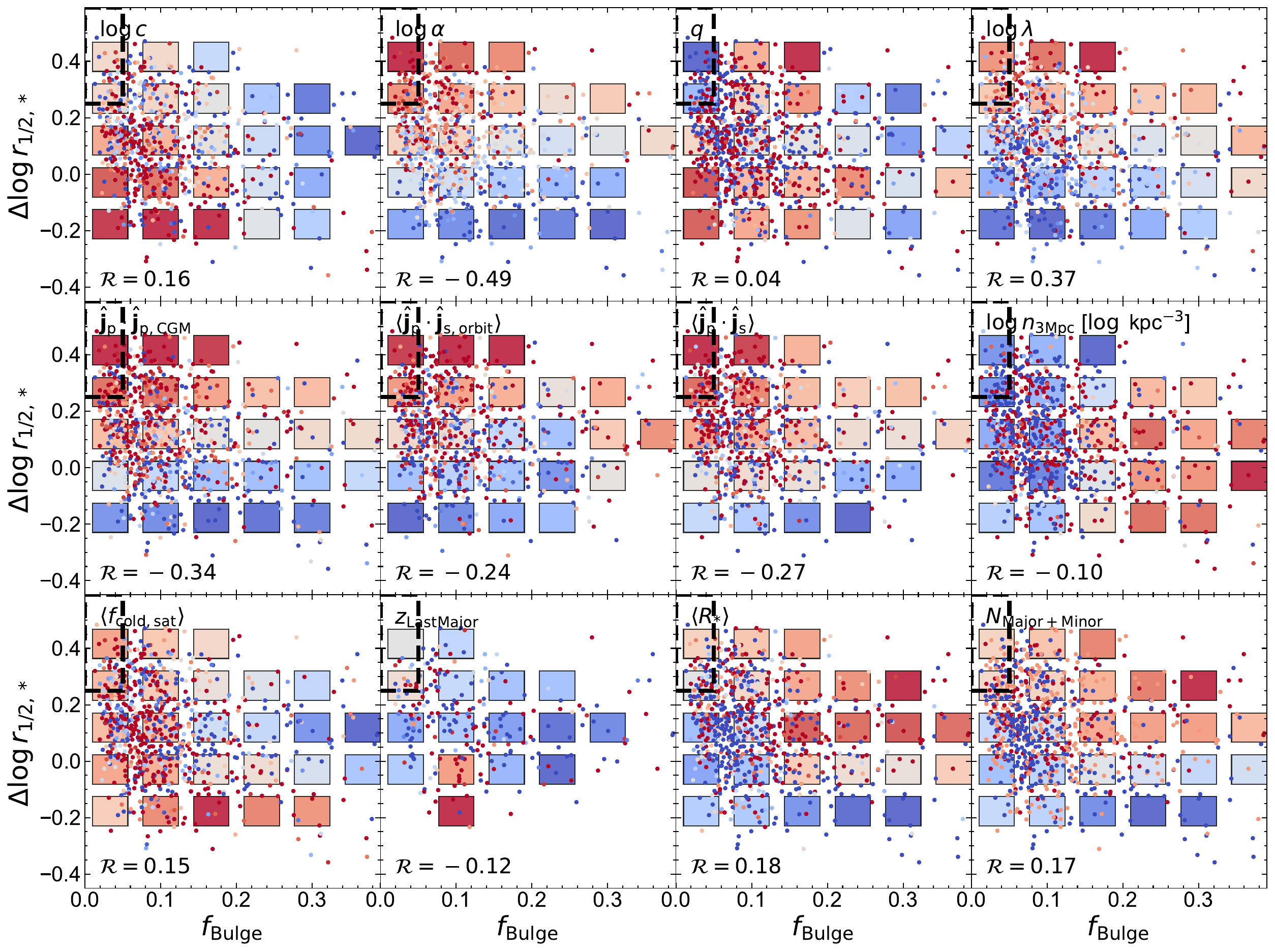}
    \caption{Similar to \Fig{rh_fdisk_X_z1}, but for the $\Delta \log r_{\rm 1/2}$ - $f_{\rm bulge}$ space.}  
    \label{fig:rh_fbulge_X_z1}
\end{figure*}

To identify the key factors for GBD formation, we examine the distributions of relevant factors in the $\Delta \log r_{\rm 1/2}$ - $f_{\rm disk}$ plane and the $\Delta \log r_{\rm 1/2}$ - $f_{\rm bulge}$ plane, 
shown in \Figs{rh_fdisk_X_z3} to \ref{fig:rh_fbulge_X_z1}. 
We mainly focus on $\Delta \log r_{\rm 1/2}$ - $f_{\rm disk}$, since $\Delta \log r_{\rm 1/2}$ - $f_{\rm bulge}$ yields basically the same qualitative conclusion.  
Here, galaxy size $\Delta \log r_{\rm 1/2}$ is defined as the offset of the logarithmic half-stellar-mass radius with respect to the median value (for galaxies with the same stellar mass at the same redshift). 
To assess the correlation between morphology and factor $X$, we compute the three-dimensional correlation coefficient, defined as
\be
\mathcal{R}^2=\frac{\mathcal{R}^2_{X,r_{1/2}}+\mathcal{R}^2_{X,f}-2\mathcal{R}_{X,r_{1/2}}\mathcal{R}_{X,f}\mathcal{R}_{r_{1/2},f}}{1-\mathcal{R}^2_{r_{1/2},f}}
\ee
where $f$ can be disk fraction $f_{\rm disk}$ or bulge fraction $f_{\rm bulge}$, and $\mathcal{R}_{i,j}$ is the Pearson correlation coefficient between quantity $i$ and quantity $j$. To better evaluate whether $X$ is correlated or anti-correlated with both size and mass fraction, or correlated with one and anti-correlated with the other, we multiply the correlation coefficient $\mathcal{R}$ by $\text{Sign}(\mathcal{R}_{X,r_{1/2}}\mathcal{R}_{X,f})$. Positive values indicate that $X$ is correlated with both size and mass fraction, while negative values mean that $X$ is correlated with one and anti-correlated with the other.

From \Fig{rh_fdisk_X_z3} to \Fig{rh_fdisk_X_z1}, we examine the $\Delta \log r_{\rm 1/2}$ - $f_{\rm disk}$ plane. 
We highlight the region with $f_{\rm disk}\geq 0.8$ and $\Delta \log r_{1/2}\geq 0.25$, which approximately corresponds to GBDs. 
If a quantity $X$ is key to GBD formation, its highest (or lowest) values are expected to correspond to large $\Delta \log r_{\rm 1/2}$ and high  $f_{\rm disk}$, rendering a diagonal trend. 

In the first row, we examine the {\it dark-halo properties}. 
The halo concentration parameter $c$ shows anti-diagonal trend, with lower values held by galaxies with larger size but smaller disk fraction. 
The trend of the Einasto-profile index $\alpha$ evolves with redshift: at $z=3$, it follows an anti-diagonal pattern, while at $z=2$ and $1$, the trend shifts to vertical and diagonal, respectively, with higher values found in larger galaxies. 
While the $\alpha$ trends are strong, we caution that it may be more of a baryonic effect, in the sense that more compact galaxies cause stronger halo contraction thus lower $\alpha$.
We thus caution against interpreting high $\alpha$ as a cause for GBD formation. 
The axis-ratio $q$ shows somewhat random distributions. However, at $z=2$ and 3, the GBD regime has large $q$ values.   
Finally, the spin parameter $\lambda$ shows basically vertical trends, with larger values for galaxies of larger sizes.

In the second row, we examine the {\it environmental factors} in terms of the alignment of CGM, the alignment of mergers, and large-scale number density. 
The three alignment angles exhibit clear diagonal trends across redshifts, with higher cosine values (better alignment) found in galaxies with larger sizes and higher disk fractions. 
In the last panel of the second row, we show the trend of the number density of neighbours within 3 Mpc, $n_{\rm 3Mpc}$. 
While the correlation is somewhat weak and the direction of the overall trend is not as clear, the GBD regime has some of the lowest number densities. 
As discussed in the paper, high-$z$ massive systems all populate high-density peaks, so the lower $n_{\rm 3Mpc}$ for GBDs simply indicates that they reside in proto-clusters or cosmic knots forming in action. 

In the last row, we examine {\it merger statistics}. 
Galaxies that experienced mergers that are more cold-gas rich tend to have higher disk fractions. 
Similarly, galaxies with quiescent merger histories also have higher disk fractions. 
However, no clear diagonal trend is observed.

Overall, coherent CGM and mergers show the most clear diagonal trend in the $\Delta \log r_{\rm 1/2}$ - $f_{\rm disk}$ space and are thus likely the most important factors for GBD formation. 
Dark-matter halo structures, especially lower concentration $c$ and higher spin $\lambda$, also play an important role in driving disks large, but do not necessarily prevent bulge formation. 
High cold-gas fraction in mergers and quiescent merger histories contribute to high disk fractions.

\clearpage

\bmhead{Acknowledgements}
 F.J. acknowledges the organizers of the 2024 Santa Cruz Galaxy Workshop, which facilitated this collaboration. 

\bmhead{Author contributions}
F.J. shaped the science goals, devised the analysis plans, and wrote the manuscript, with substantial input from  S.C., W.W., L.C.H., Y.P., and J.W.. 
J.L. carried out the simulation analysis, with input from F.J. on halo-related measurements and X.S. on radiative transfer with {\tt SKIRT}.
B.J. performed surface photometry using {\tt GALFIT} with guidance from L.C.H., and Z.G. carried out SED fitting with guidance from Y.P., using the mock observations generated by J.L., with substantial technical input from W.W..

\bmhead{Competing interests}
The authors declare that they have no competing interests.

\bmhead{Data availability}
The IllustrisTNG simulations are publicly available at \href{www.tng-project.org/data}{https://www.tng-project.org/data}. Additional data directly related to this publication are available on request from the corresponding authors.


\bibliography{GBD}

\end{document}